\documentclass[12pt]{article}

\usepackage{tikz-cd}
\usepackage{adjustbox}
\usepackage{capt-of}
\usepackage[utf8]{inputenc}
\usepackage{textcomp}
\usepackage{amsmath}
\usepackage[shortlabels]{enumitem}
\usepackage{chetnew}
\usepackage{tikz}
\usepackage{amssymb,amsfonts,mathrsfs,dsfont,yfonts,bbm}\usetikzlibrary{calc}
\usepackage{xcolor}
\usepackage{braket}
\usepackage[normalem]{ulem}
\usepackage{import}
\usepackage{cancel}
\usepackage{booktabs}
\usepackage{varioref}
\usepackage{verbatim}
\usepackage{hyperref}
\usepackage{graphicx}
\usepackage{bm}
\usepackage{tikz-cd}
\usepackage[cmtip,all]{xy}
\newcommand{\longsquiggly}{\xymatrix{{}\ar@{~>}[r]&{}}}

\newcommand{\cI}{\mathcal{I}}

\newcommand{\cC}{\mathcal{C}}
\newcommand{\sO}{\mathscr O}

\newcommand{\beqn}{\begin{eqnarray}}
\newcommand{\eeqn}{\end{eqnarray}}

\newcommand{\be}{\begin{equation}}
\newcommand{\ee}{\end{equation}}
\newcommand{\bea}{\begin{eqnarray}}
\newcommand{\eea}{\end{eqnarray}}

\newcommand{\p}{\partial}
\newcommand{\CL}{\mathcal{L}}

\newcommand{\CD}{\mathcal{D}}
\newcommand{\CE}{\mathcal{E}}

\newcommand{\CB}{\mathcal{B}}
\newcommand{\CC}{\mathcal{C}}

\newcommand{\CO}{\mathcal{O}}

\newcommand{\CI}{\mathcal{I}}
\newcommand{\CN}{\mathcal{N}}

\newcommand*{\boxcoloro}{orange}
\makeatletter
\newcommand{\boxedo}[1]{\textcolor{\boxcoloro}{%
\tikz[baseline={([yshift=-1ex]current bounding box.center)}] \node [rectangle, minimum width=1ex,rounded corners,draw] {\normalcolor\m@th$\displaystyle#1$};}}
 \makeatother
\newcommand*{\boxcolorr}{red}
\makeatletter
\newcommand{\boxedr}[1]{\textcolor{\boxcolorr}{%
\tikz[baseline={([yshift=-1ex]current bounding box.center)}] \node [rectangle, minimum width=1ex,rounded corners,draw] {\normalcolor\m@th$\displaystyle#1$};}}
 \makeatother
\newcommand*{\boxcolorb}{blue}
\makeatletter
\newcommand{\boxedb}[1]{\textcolor{\boxcolorb}{%
\tikz[baseline={([yshift=-1ex]current bounding box.center)}] \node [rectangle, minimum width=1ex,rounded corners,draw] {\normalcolor\m@th$\displaystyle#1$};}}
 \makeatother
\newcommand*{\boxcolorg}{green}
\makeatletter
\newcommand{\boxedg}[1]{\textcolor{\boxcolorg}{%
\tikz[baseline={([yshift=-1ex]current bounding box.center)}] \node [rectangle, minimum width=1ex,rounded corners,draw] {\normalcolor\m@th$\displaystyle#1$};}}
 \makeatother
 \newcommand*{\boxcolorp}{purple}
\makeatletter
\newcommand{\boxedp}[1]{\textcolor{\boxcolorp}{%
\tikz[baseline={([yshift=-1ex]current bounding box.center)}] \node [rectangle, minimum width=1ex,rounded corners,draw] {\normalcolor\m@th$\displaystyle#1$};}}
 \makeatother
  \newcommand*{\boxcolorc}{cyan}
\makeatletter
\newcommand{\boxedc}[1]{\textcolor{\boxcolorc}{%
\tikz[baseline={([yshift=-1ex]current bounding box.center)}] \node [rectangle, minimum width=1ex,rounded corners,draw] {\normalcolor\m@th$\displaystyle#1$};}}
 \makeatother
  \newcommand*{\boxcolory}{yellow}
\makeatletter
\newcommand{\boxedy}[1]{\textcolor{\boxcolory}{%
\tikz[baseline={([yshift=-1ex]current bounding box.center)}] \node [rectangle, minimum width=1ex,rounded corners,draw] {\normalcolor\m@th$\displaystyle#1$};}}
 \makeatother

\usepackage{amsmath}	

\begin{document}
\preprint{QMUL-PH-22-35}
 
\title{Argyres-Douglas Avatars of \\[3mm] Coulomb Branch Physics}

\author{Chinmaya Bhargava,$^{\spadesuit,1}$ Matthew Buican,$^{\diamondsuit,1,2,3}$ and Hongliang Jiang$^{\clubsuit,1}$}

\affiliation{\smallskip $^1$CTP and Department of Physics and Astronomy\\
Queen Mary University of London, London E1 4NS, UK\\[2mm] $^2$Simons Center for Geometry and Physics, SUNY, Stony Brook, NY 11794, USA\\[2mm] $^3$School of Natural Sciences, Institute for Advanced Study, Princeton, NJ 08540, USA\emails{$^{\spadesuit}$c.bhargava@qmul.ac.uk, $^{\diamondsuit}$m.buican@qmul.ac.uk, $^{\clubsuit}$ h.jiang@qmul.ac.uk}}

\abstract{We study ultraviolet (UV) incarnations of deep infrared (IR) physics on the Coulomb branch of the simplest interacting 4D $\CN=2$ superconformal field theory: the minimal Argyres-Douglas (MAD) theory. One of the most basic properties of the Coulomb branch is an emergent infinite-dimensional higher-spin symmetry. While the MAD theory is interacting and therefore does not have such a symmetry, we find UV operators that encode the emergent complex higher-spin symmetry on the Coulomb branch. Moreover, we show that cousins of these UV operators give rise to cousins of the IR higher-spin multiplets. In terms of superconformal representation theory, we are led to a conjecture on the exact spectrum of $\bar\CC_{R,r(j,\bar j)}$ multiplets in the MAD theory for all $R$, $r$, $j$, and $\bar j$ satisfying $R+\bar j -j+1=0$, thereby making progress towards a full characterization of the protected spectrum.  Along the way, we give a geometrical interpretation of these operators and include them in an extension of the Coulomb branch / $\CN=2$ chiral operator correspondence.}
\bigskip

\date{November 2022}

\setcounter{tocdepth}{2}
\maketitle
\toc

\newsec{Introduction}
One of the most important and difficult open problems in theoretical physics is to exactly solve an interacting quantum field theory (QFT) in $d>2$ spacetime dimensions. Such a QFT will likely be highly supersymmetric, and the solution will probably involve an interplay between ultraviolet (UV) interactions and a moduli space of vacua in the infrared (IR). In the context of $d=4$, the most universal moduli space of vacua for an interacting theory with extended supersymmetry is the Coulomb branch, where various non-perturbative results have been extracted, starting with the classic works of Seiberg and Witten \cite{Seiberg:1994rs,Seiberg:1994aj}.\footnote{In the case of 4D $\CN=4$ super-Yang-Mills theory (and various close cousins), integrability, holography, and techniques from scattering amplitudes have been used to extract highly non-trivial parts of the spectrum and dynamics (e.g., see \cite{Beisert:2010jr,Gadde:2010zi,Arkani-Hamed:2022rwr} and references therein). Our approach here, by contrast, will be to focus on an even simpler (albeit less immediately accessible) {\it isolated} interacting theory.}

One of the most basic properties of the Coulomb branch is the existence of an emergent infinite-dimensional higher-spin symmetry in the deep IR. The emergence of this symmetry goes hand-in-hand with our ability to solve the theory in this regime.

While UV interactions along a renormalization group (RG) flow to the Coulomb branch spoil the higher-spin symmetry at non-zero energies \cite{Coleman:1967ad,Haag:1974qh,Maldacena:2011jn,Alba:2015upa}, it is natural to wonder if one can \lq\lq see" aspects of the higher spin symmetries in the UV operator content. Understanding this spectral   imprint may then help in solving the theory at all energy scales.\footnote{These ideas are similar in spirit to the question of studying continuum limits of integrable lattice models (e.g., see the recent discussion and references in \cite{Aasen:2020jwb}).}

In this paper, we consider the example of the minimal Argyres-Douglas (MAD) theory \cite{Argyres:1995jj,Argyres:1995xn} and the flow onto its one-complex-dimensional Coulomb branch (e.g., see \cite{Argyres:2015ffa}). There are various reasons to consider this system when studying emergence of higher-spin symmetry and the resulting Noether currents:
\begin{itemize}
\item From the perspective of effective field theory, the MAD SCFT is particularly simple: it can be understood as an interacting SCFT that emerges at special points on the Coulomb branch of certain low-rank $\CN=2$ gauge theories \cite{Argyres:1995jj,Argyres:1995xn}. Close to these points, the MAD theory can roughly be thought of as an abelian gauge multiplet coupled to two massless hypermultiplets with mutually non-local electromagnetic charges. These massless degrees of freedom therefore constitute the \lq\lq simplest" possible ingredients for an interacting $\CN=2$ theory in 4D.
\item More abstractly, certain sectors of the MAD operator algebra are as simple as possible for a unitary theory with a Coulomb branch \cite{Buican:2021elx,Bhargava:2022cuf}.
\item Infinitely many OPE coefficients agree between the MAD theory and the free $\CN=2$ Maxwell theory (i.e., the theory at generic points on the Coulomb branch) \cite{Buican:2021elx}.
\end{itemize}
Therefore, given the simplicity of the MAD SCFT and its \lq\lq closeness" to the Coulomb branch theory, it is natural to study the UV origins of higher spin symmetry in this system.

The higher-spin symmetries of the theory at generic points on the Coulomb branch can be organized into three infinite sets: two complex sets of higher-spin current multiplets that are conjugates of each other, along with a set of real higher-spin current multiplets (see section \ref{coulomb} for details). In this note we focus on the complex current multiplets, the simplest of which are housed in so-called $\hat\CC_{0(1,0)}$ multiplets\footnote{We mainly follow the conventions of \cite{Dolan:2002zh} (up to the sign of $U(1)_r$ charge). See also \cite{Dobrev:1985qv,Cordova:2016emh} for related discussions in different conventions.} whose superconformal primary (SCP) takes the form  
\begin{equation}\label{HSprim1} 
 \CO _{\alpha\beta}:=\epsilon_{ij}\lambda^i_{\alpha}\lambda^j_{\beta}+\kappa\phi F_{\alpha\beta}\in\hat\CC_{0(1,0)}~.
\end{equation}
Here $\lambda^i_{\alpha}$ are the gauginos ($i=1,2$ is an $SU(2)_R$ index, while $\alpha,\beta=\pm$ are spinor indices), $\phi$ is the scalar primary of the vector multiplet, $F_{\alpha\beta}$ is the field strength, and $\kappa$ is a constant that can be determined by demanding that $ \CO _{\alpha\beta}$ is an SCP. The higher-spin currents are superconformal descendants of this operator, and their conservation equations correspond to the simultaneous shortening conditions (along with those obtained by $SU(2)_R$ lowering)
\begin{equation}\label{HScond}
Q^{1\alpha} \CO _{\alpha\beta}=0~,\ \ \ (\bar Q^1)^2 \CO _{\alpha\beta}=0~.
\end{equation}

To understand what happens to the $\hat\CC_{0(1,0)}$ multiplet when we flow back up to the UV, let us, for simplicity, suppose we are near a point on the MAD Coulomb branch where a charged hypermultiplet couples (marginally irrelevantly) to the gauge multiplet through the superpotential term (as described in \cite{Argyres:2015ffa}, we should turn on a VEV and a relevant prepotential deformation in the UV theory to reach this point)\footnote{We thank Z.~Komargodski for a discussion of these points.}
\begin{equation}\label{Wdef}
W=g\phi q\tilde q~,
\end{equation}
where $q$, $\tilde q$ are the hypermultiplet fields, and $g$ is the gauge coupling. In the presence of this coupling, it is straightforward to check that the second equation in \eqref{HScond} is still obeyed\footnote{This statement follows from the fact that $\bar Q^1_{\dot\alpha}\phi=0$ and so $\bar Q^1_{\dot\alpha}\lambda^1_{\beta}=(\bar Q^1)^2\lambda^2_{\alpha}=(\bar Q^1)^2F_{\alpha\beta}=0$. }
\begin{equation}\label{barC}
(\bar Q^1)^2 \CO _{\alpha\beta}=0~,
\end{equation}
but the first equation is not\footnote{This fact can be understood from the observation that $Q^{1\alpha}F_{\alpha\beta}\sim Q^2_{\beta}F\sim \bar gQ^2_{\beta}(\bar q\bar{\tilde q})$ is the only term in $Q^{1\alpha} \CO _{\alpha\beta}$ that contains hypermultiplet fermions. Note that in this derivation, $F=-\partial_{\bar\phi}\overline W$ is a component of the auxiliary field of the free vector.}
\begin{equation}\label{HSviolated}
Q^{1\alpha} \CO _{\alpha\beta}\ne0~.
\end{equation}

Physically, \eqref{HSviolated} is a consequence of the fact that interactions break higher-spin symmetry. More interestingly for us in what follows, a multiplet satisfying the preserved equation \eqref{barC} obeys the shortening conditions of a so-called $\bar\CC$ multiplet
\begin{equation}\label{map}
 \CO  _{\alpha\beta}\in\hat\CC^{\rm\, Free}_{0(1,0)}\   \longsquiggly  \CO _{\alpha\beta}\in \bar\CC_{0,r(1,0)}~,
\end{equation}
where $r>1$ is a $U(1)_r$ charge.

So far, our discussion has been in the deep IR effective theory of the MAD Coulomb branch. However, we are more interested in what happens to the $\bar\CC_{0,r(1,0)}$ multiplet as we go into the deep UV and reach the interacting SCFT. Intuitively, we expect that the map in \eqref{map} (appropriately defined) is robust and that we need only determine $r$ (which is broken in the flow to the IR). One reason is that higher spin symmetry is sensitive only to interactions and not to the breaking of conformal symmetry (the theory of a free massive field still has higher-spin symmetry). More technically, we expect that corrections to the operator \eqref{HSprim1} involve the prepotential and other holomorphic quantities.

In fact, we argue that the map in \eqref{map} extends to the MAD theory itself if we set $r=7/5$. Reversing the order of UV and IR in \eqref{map} we have a one-to-one correspondence\footnote{In section \ref{MADfvmap}, we will explain a precise sense in which the operator on the LHS of \eqref{HSmap} flows to the operator on the RHS.}
\begin{equation}
\bar\CC_{0,7/5(1,0)}^{\rm\, MAD}\ \longrightarrow\ \hat\CC^{\rm \, Free}_{0(1,0)}~.
\end{equation}
We can then repeat the above logic for all complex higher-spin symmetry multiplets on the Coulomb branch. For any of these symmetries, we find the following one-to-one correspondence
\begin{equation}\label{HSmap0}
\bar\CC_{0,{7\over5}(k,k-1)}^{\rm\, MAD}\ \longrightarrow\ \hat\CC^{\rm \, Free}_{{0}(k,k-1)}~,\ \ \ k\in\mathbb{Z}_{\ge1}~.
\end{equation}

But the story above is just a \lq\lq small" (albeit infinite) piece of a larger correspondence. A unifying property of the IR operators involved is that they appear as certain non-singular (but non-chiral\footnote{Here we have in mind \lq\lq non-chiral" in both the $\CN=2$ and $\CN=1$ senses of the word. See appendix A of \cite{Bhargava:2022cuf} for an explanation of why the multiplets on the RHS of \eqref{IROPE} are non-chiral.}) terms in the following $(n+p+1)$-fold OPEs\footnote{The two-fold OPE is the usual OPE. The three-fold OPE is the OPE of $\bar\CD_{0(0,0)}$ with all terms appearing on the RHS of the two-fold OPE. One can then define the $N$-fold OPE, for all $N>3$, by recursion.}
\begin{equation}\label{IROPE}
\bar\CD_{0(0,0)}^{\times(n+p+1)}\ni \bar\CC^{\rm \, Free}_{{p\over2},n+{p\over2}(k,k-1-{p\over2})}~,\ \hat\CC^{\rm \, Free}_{{p\over2}(k,k-1-{p\over2})}~,\ \ \ n\in\mathbb{Z}_{\ge1}~,\ \ \ p\in\mathbb{Z}_{\ge0}~,
\end{equation}
where the higher-spin currents sit in $\hat\CC_{{p\over2}(k,k-1-{p\over2})}$ multiplets for $p=0$, and $\bar\CD_{0(0,0)}$ is the chiral part of the IR vector multiplet (i.e., its SCP is $\phi$ and is annihilated by all the $\bar Q^i_{\dot\alpha}$). In the UV, the $\bar\CD_{0(0,0)}$ multiplet descends from the well-known dimension $6/5$ $\CN=2$ chiral multiplet of the MAD theory,\footnote{The scaling dimension of the primary can be read off from the Seiberg-Witten solution.} $\bar\CE_{6/5}$ (like $\phi$, the corresponding SCP is annihilated by all the $\bar Q^i_{\dot\alpha}$). In terms of these degrees of freedom, we argue that the UV version of \eqref{IROPE} is
\begin{equation}\label{UVOPE}
\bar\CE^{\times(n+p+1)}_{6/5}\ni\bar\CC^{\rm\, MAD}_{{p\over2},{1\over10}(2+12n+7p)(k,k-1-{p\over2})}~,
\ \ \ n\in\mathbb{Z}_{\ge1}~,\ \ \ p\in\mathbb{Z}_{\ge0}~.
\end{equation}
Note that the operators appearing on the RHS of \eqref{IROPE} and \eqref{UVOPE} satisfy the following relation
\begin{equation}\label{scond}
s_{ \CO }:=R_{ \CO }+\bar j_{ \CO }-j_{ \CO }=-1~, 
\end{equation}
where $ \CO $ is the SCP of the multiplet in question.\footnote{There are also $s_{ \CO }>-1$ $\bar\CC$ multiplets appearing in the OPEs described in \eqref{IROPE} and \eqref{UVOPE}, but we leave an analysis of these additional multiplets for future work \cite{BBJ2}.} The subset of operators in \eqref{UVOPE} with $p=0$ and $k=1$ appeared in \cite{Bhargava:2022cuf} (along with an infinite family of $s_{ \CO }=0$ multiplets whose generalization we do not study here).

As we will discuss below, the operators appearing in the UV and IR OPEs are subject to the following one-to-one relations
\begin{equation}\label{CCmap0}
\bar\CC_{{p\over2},{1\over10}(2+12n+7p)(k,k-1-{p\over2})}^{\rm\, MAD}\ \longrightarrow\ \ \bar\CC^{ \rm\, Free}_{{p\over2},n+{p\over2}(k,k-1-{p\over2})}~,\ \ \ n\in\mathbb{Z}_{>1}~,
\end{equation} 
and, for $n=1$
\begin{equation}\label{CChatmap0}
\bar\CC_{{p\over2},{7\over10}(2+p)(k,k-1-{p\over2})}^{\rm\, MAD}\ \longrightarrow\ \ \hat\CC^{ \rm\, Free}_{{p\over2}(k,k-1-{p\over2})}~.
\end{equation}
The map involving IR higher spin currents in \eqref{HSmap0} is then the special case corresponding to $p=0$.

Purely from the IR perspective, inspired by the logic below \eqref{HScond} and its generalization for all complex higher-spin multiplets, combined with the explicit form of the $\bar\CC$ and $\hat\CC$ multiplets alluded to in \eqref{IROPE} (see section \ref{coulomb} for details), it is tempting to suggest that (up to $U(1)_r$ charge assignments) the map in \eqref{CCmap0} and \eqref{CChatmap0} holds for all $p$ and $n$. In particular, we are motivated to find $\bar\CC$ UV multiplets with the matching $R$, $j$ and $\bar j$ quantum numbers and to ignore other multiplets. Moreover, since the operators on the RHS of \eqref{CCmap0} and \eqref{CChatmap0} appear as non-singular terms of the OPEs in \eqref{IROPE}, it is natural to use the UV/IR $\CN=2$ chiral multiplet map to construct the UV operators on the LHS of \eqref{CCmap0} and \eqref{CChatmap0}. In section \ref{MADfvmap} we will describe this map in more detail and see that it is consistent, in a highly non-trivial fashion, with the MAD superconformal index.

A remaining question is to understand if there can be additional UV multiplets of the type appearing in \eqref{CCmap0} and \eqref{CChatmap0} that are not detected by the index and that flow to zero in the IR.\footnote{A priori, one might imagine that such additional multiplets could also flow to long multiplets in the IR. However, we can rule out this possibility by starting in the MAD theory and turning on a vev of an $\CN=2$ chiral operator to spontaneously break $U(1)_r$ and flow onto the Coulomb branch while preserving $\CN=2$ SUSY. In this case, all the vanishing supercharge actions on operators in a UV $\bar\CC$ multiplet are preserved. This observation implies that the $\bar\CC$ multiplets cannot lengthen in the IR (however, they are allowed to shorten; indeed, precisely this situation occurs in \eqref{CChatmap0}).} We conjecture that such UV operator decoupling does not occur, and we give the following non-trivial pieces of evidence to substantiate this claim:
\begin{itemize}
\item As is well-known from the Seiberg-Witten construction, no $\bar\CE$ operators in the MAD theory flow to zero. Similarly, the results of \cite{Buican:2021elx} strongly suggest that $\hat\CC$ multiplets in the MAD theory flow to $\hat\CC$ multiplets in the IR (note that $\hat\CC$  multiplets in the MAD theory always have the same left and right spins, so they cannot  appear in \eqref{CCmap0} and \eqref{CChatmap0}).\footnote{ This latter statement is highly non-trivial and is typically violated in $\CN=2$ SCFTs. For example, consider the case of the simplest $\CN=2$ SCFT with $SU(2)$ flavor symmetry, the so-called $(A_1,A_3)$ theory originally constructed in \cite{Argyres:1995xn}. In that case, flows to the Coulomb branch decouple infinitely many $\hat\CC$ (and $\hat\CB$) multiplets. By the results of \cite{Buican:2021elx}, this statement holds more generally anytime a theory has a non-trivial Higgs branch. \label{Higgstheory}} Moreover, in \cite{Bhargava:2022cuf} we used the $\CN=1\to\CN=2$ RG flow of \cite{Maruyoshi:2016tqk} to prove that none of the $\bar\CC_{0,r(j,0)}$ multiplets flow to zero. This set of multiplets includes the $\bar\CC_{0,r(1,0)}$ multiplets that form a subset (with $p=0$ and $k=1$) of the multiplets appearing in \eqref{CCmap0} and \eqref{CChatmap0}. Our present claim extends these results to the broader set of related multiplets with $p\ge0$ and $k\ge1$.
\item In \cite{Buican:2021elx} and \cite{Bhargava:2022cuf} we saw that the $\bar\CE$ sector, the $\hat\CC$ sector, and the parts of the $\bar\CC$ sector we were able to study were as simple as possible for a unitary theory with a Coulomb branch. Our above discussion extends these statements to the operators in \eqref{CCmap0} and \eqref{CChatmap0} (which generalize a subset of operators studied in \cite{Bhargava:2022cuf} and described in the previous bullet).
\item In \cite{Bhargava:2022cuf}, we found highly non-trivial evidence that the local operator algebra of the MAD theory is generated by the $(n,m)$-fold $\bar\CE_{6/5}^{\times n}\times\CE_{-6/5}^{\times m}$ OPEs.\footnote{Recall that the energy momentum tensor multiplet appears in the $\bar\CE_{6/5}\times\CE_{-6/5}$ OPE (see footnote 34 of \cite{Bhargava:2022cuf}).}  Our results here are compatible with and reinforce this conjecture. For the case of multiplets with the quantum numbers of those on the RHS of \eqref{CCmap0} and \eqref{CChatmap0}, it is natural that they must appear in non-singular pieces of the OPEs in \eqref{UVOPE} (otherwise, we would expect an inconsistency with IR OPEs). As we will see below, our construction gives the maximal number of such operators and so there is no room for additional multiplets to appear.
\item As we will see in section \ref{MADfvmap}, by fixing the mapping of two degrees of freedom, we can explain the structure of the $s_{ \CO }=-1$ contributions of the MAD superconformal index. The simplicity and predictiveness of this map suggests that our construction is correct. Although the index can in principle suffer from cancellations, we will see that operators of the form on the LHS of \eqref{CCmap0} and \eqref{CChatmap0} cannot cancel amongst themselves in the UV index (note that the operators on the RHS do cancel in the index of the IR Coulomb branch theory). This fact is a confirmation of the robustness of our approach.
\end{itemize}

From the above discussion, and using the fact that the operators on the RHS of \eqref{CCmap0} and \eqref{CChatmap0} are the only operators in the IR $\bar\CC$ multiplets satisfying \eqref{scond}, we are led to the following central claim / conjecture:

\bigskip
\noindent
{\bf Claim:} The spectrum of $\bar\CC_{{p\over2},r(k,k-1-{p\over2})}$ multiplets in the MAD theory is given by
\begin{eqnarray}\label{mainresult}
N_{\bar\CC^{\rm\, MAD}_{{p\over2},r(k,k-1-{p\over2})}}=\begin{cases}
\sum_{i=0}^p(-1)^i\Big [N\left(p-i,n+i,k-{i\over2}\right)-N\left(p-i,n+i,k-{i\over2}-{1\over2}\right)\Big ]~,\ r=r(n,p)~,\\ 
0~, \ \text{otherwise}~,
\end{cases}
\end{eqnarray}
where
\begin{eqnarray}
N(p,n,k)&:=&\sum_{ q =1}^{ q _{\rm max}}  \sum_{\ell =0}^{2k- q -p} (-1)^{ q +1} R(\ell, q +p)S(2k- q -p-\ell ,n- q +1)~,\cr r(n,p)&:=&{1\over10}(2+12n+7p) ~,\ \ \ n\in\mathbb{Z}_{\ge1}~,\ \ \ p\in\mathbb{Z}_{\ge0}~.
\end{eqnarray}
Here $S(n,k)$ is the number of partitions of $n$ into $k$ non-negative integers, and $R(n,k)$ is the number of partitions of $n$ into $k$ {\it distinct} non-negative integers.  See Appendix \ref{PartInt} for details on these partitions and their generating functions.
Finally, we define $ q _{\rm max}$ to be the largest $ q $ such that $R(2k- q -p, q +p)\ne0$. By CPT, the multiplicities of the conjugate $\CC_{{p\over2},-r(k-1-{p\over2},k)}$ multiplets are also given by \eqref{mainresult}.

\bigskip
\noindent
Indeed, we will see in section \ref{coulomb} that the RHS of the above equations match the multiplicities in the IR theory under the maps in \eqref{CCmap0} and \eqref{CChatmap0}.

We conclude by briefly summarizing the rest of the paper. In the next section, we develop a theory of Coulomb branch counting that allows us to compute the spectrum of IR multiplets in \eqref{CCmap0} and \eqref{CChatmap0}. To arrive at these results, we develop a relevant notion of cohomology and prove that it is trivial. This picture also gives a geometrical construction for the IR multiplets of interest. Then, in section \ref{MADfvmap}, we extend these results to the MAD theory and check compatibility with the superconformal index. We conclude with a series of open questions and avenues for future work. Finally, we include two appendices with additional details and perspectives.

\newsec{The free vector:  the deep IR of the Coulomb branch}\label{coulomb}
 In this section we begin by discussing the complex higher-spin symmetry multiplets in \eqref{HSmap0} that are present at generic points in the deep IR of the MAD Coulomb branch. We then generalize these operators and give the full spectrum of IR multiplets appearing on the RHS of \eqref{CCmap0} and \eqref{CChatmap0}. In order to complete this spectral analysis, we introduce a notion of cohomology and solve it in subsection \ref{cohomology}. As we will see, this discussion also leads to a geometrical construction of the operators of interest.

\subsection{Higher-spin symmetry on the Coulomb branch}
The complex higher-spin symmetry multiplets house corresponding Noether currents and therefore must be quadratic in the Maxwell fields. On general grounds, higher-spin current multiplets are of type $\hat\CC_{0(j, \bar j)}$ (e.g., see \cite{Beem:2013sza}). These are examples of so-called \lq\lq Schur" multiplets housing \lq\lq Schur" operators sitting as (conformal primary) level-two superconformal $Q^1_+\bar Q^1_{\dot+}$ descendants with highest $SU(2)_R$ and Lorentz weight \cite{Gadde:2011uv,Beem:2013sza}. In the case of the free vector theory, the Schur operators are generated as a ring by $\lambda^1_+$, $\bar\lambda^1_{\dot+}$, and $\partial_{+\dot+}$.

Since we are interested in complex higher-spin current multiplets, we should build the Schur operators in our multiplets of interest from two $\lambda^1_+$'s and arbitrarily many $\partial_{+\dot+}$'s (we also obtain conjugate multiplets via $\lambda^1_+\leftrightarrow\bar\lambda^1_{\dot+}$). In particular, we can associate each possible Schur operator with a pair of integers, $(n,m)$, as follows
\begin{equation}
(n,m)\longleftrightarrow\partial_{+\dot+}^n\lambda^1_+\partial^m_{+\dot+}\lambda^1_+~,\ \ \ n+m=K~,
\end{equation}
where, without loss of generality, we may take $n<m$ (for $n=m$, the operator vanishes by Fermi statistics). If $K$ is odd, then there are $\lfloor K/2 \rfloor+1=(K+1)/2$ such operators. When $K$ is even there are $K/2$ such operators. The number of primaries with $K$ odd is then $(K+1)/2-(K-1)/2=1$, while the number of primaries with $K$ even is $K/2-K/2=0$. Therefore, we see that the spectrum of complex higher-spin multiplets is\footnote{By similar logic, for the complex conjugate multiplets obtained by interchanging $\lambda^1_+\leftrightarrow\bar\lambda^1_{\dot+}$, we have
\begin{eqnarray}\label{NumHSconj}
N_{\hat\CC_{0(k-1,k)}}=\begin{cases}
1~,\ k=(K+1)/2\in\mathbb{Z}_{>0}~,\\ 
0~, \ \text{otherwise}~.
\end{cases}~.
\end{eqnarray}
Although we won't discuss them here, there are also real higher-spin multiplets involving Schur operators of the form $\partial_{+\dot+}^n\lambda^1_+\partial^m_{+\dot+}\bar\lambda^1_{\dot+}$. One can check that the spectrum of these multiplets is
\begin{eqnarray}\label{NumHSreal}
N_{\hat\CC_{0(k,k)}}=1~,\ \ \   k=\frac12, 1, \frac32, 2, \cdots~.
\end{eqnarray}
}
\begin{eqnarray}\label{NumHS}
N_{\hat\CC_{0(k ,k-1)}}=\begin{cases}
1~,\ k=(K+1)/2\in\mathbb{Z}_{>0}~,\\ 
0~, \ \text{otherwise}~.
\end{cases}~.
\end{eqnarray}

Let us now consider various natural generalizations of the above higher-spin multiplets. Although these multiplets are clearly not chiral (multiplets housing chiral operators in both the $\CN=1$ and $\CN=2$ senses coincide for the free vector \cite{Bhargava:2022cuf} and have SCPs with $j=\bar j=0$), it is clear they are built from operators in the free vector chiral multiplet, $\bar\CD_{0(0,0)}$ with primary $\phi$ (and highest $SU(2)_R$ and Lorentz weight level-one descendant $\lambda^1_+$). Said differently, these multiplets appear in the OPE of chiral multiplets
\begin{equation}
\bar\CD_{0(0,0)}^{\times2}:=\bar\CD_{0(0,0)}\times\bar\CD_{0(0,0)}\ni\hat\CC_{0(k,k-1)}~.
\end{equation}

\subsection{The quasi-chiral sector}
It is then natural to consider short multiplets appearing in the $N$-fold $\bar\CD_{0(0,0)}^{\times N}$ OPEs. Such multiplets must be built from normal-ordered products of fields in $\bar\CD_{0(0,0)}$ and derivatives. In addition to the multiplets in \eqref{NumHS}, other multiplets in this sector include $\bar\CD_{0(0,0)}$ itself and the chiral $\bar\CE_{n}:=\bar\CD_{0(0,0)}^n$ multiplets for $n>1$ (here $\bar\CE_1=\bar\CD_{0(0,0)}$).\footnote{$\bar\CD_{0(0,0)}^n$ denotes taking the $n$th power of the $\bar\CD_{0(0,0)}$ primary and applying Poincar\'e supercharges to complete the multiplet. This product should be distinguished from the $n$-fold OPE, $\bar\CD_{0(0,0)}^{\times n}$.}

\begin{table}[h!]
\centering
\begin{tabular}{|c|cccc|ccccccc|}
\hline
$\delta = -2$ & $\bar{Q}^1_{\dot{+}}$ &  &  &  &                             &                               &  &  &  &  &  \\
$\delta=0$ &
  $Q^{1}_{\alpha}$ &
  $\bar{Q}^{1}_{\dot{-}}$ &
  $\bar{Q}^{2}_{\dot{+}}$ &
   &
  $\lambda^{1}_{\alpha}$ &
  $\phi$ &
  $\partial_{\alpha \dot{+}}$ &
  $\bar{F}_{\dot{+} \dot{+}}$ &
  $\bar{\lambda}^{1}_{\dot{+}}$ &
   &
   \\
$\delta=2$ &
  $Q^{2}_{\alpha}$ &
  $\bar{Q}^{2}_{\dot{-}}$ &
   &
   &
  $\lambda^{2}_{\alpha}$ &
  $F_{\alpha \beta}$ &
  $\partial_{\alpha \dot{-}}$ &
  $\bar{\phi}$ &
  $\bar{F}_{\dot{-}\dot{+}}$ &
  $\bar{\lambda}^{1}_{\dot{-}}$ &
  $\bar{\lambda}^{2}_{\dot{+}}$ \\
$\delta=4$    &                       &  &  &  & $\bar{F}_{\dot{-} \dot{-}}$ & $\bar{\lambda}^{2}_{\dot{-}}$ &  &  &  &  & \\
\hline
\end{tabular}
\caption{$\delta$ quantum numbers of the $\CN = 2$ Poincar\'e supercharges and fields in the free vector multiplet, $\bar\CD_{0(0,0)}\oplus\CD_{0(0,0)}$.}
\label{FVDeltaQDelta}
\end{table}

On general grounds, the remaining multiplets must be of type $\hat\CC_{R(j,\bar j)}$ and $\bar\CC_{R,r(j,\bar j)}$ (see \cite{Bhargava:2022cuf} for an explanation).\footnote{We also have multiplets conjugate to those we have introduced.} We would like to understand the subset of these multiplets appearing in the $N$-fold $\bar\CD_{0(0,0)}^{\times N}$ OPE. To that end, note that all $\hat\CC$ and $\bar\CC$ multiplets have highest $SU(2)_R$ and Lorentz-weight primaries, $ \CO $, satisfying
\begin{equation}
\delta_{ \CO }:=\Delta_{ \CO }-\left(2R_{ \CO }+2\bar j_{ \CO }+r_{ \CO }\right)=2~.
\end{equation}
Since these multiplets obey (semi) shortening conditions, they also contribute to the superconformal index. In fact, the index itself is only sensitive to the subspace of local operators with $\delta=0$.

\begin{figure}[h!]
\centering
\includegraphics[width=\textwidth]{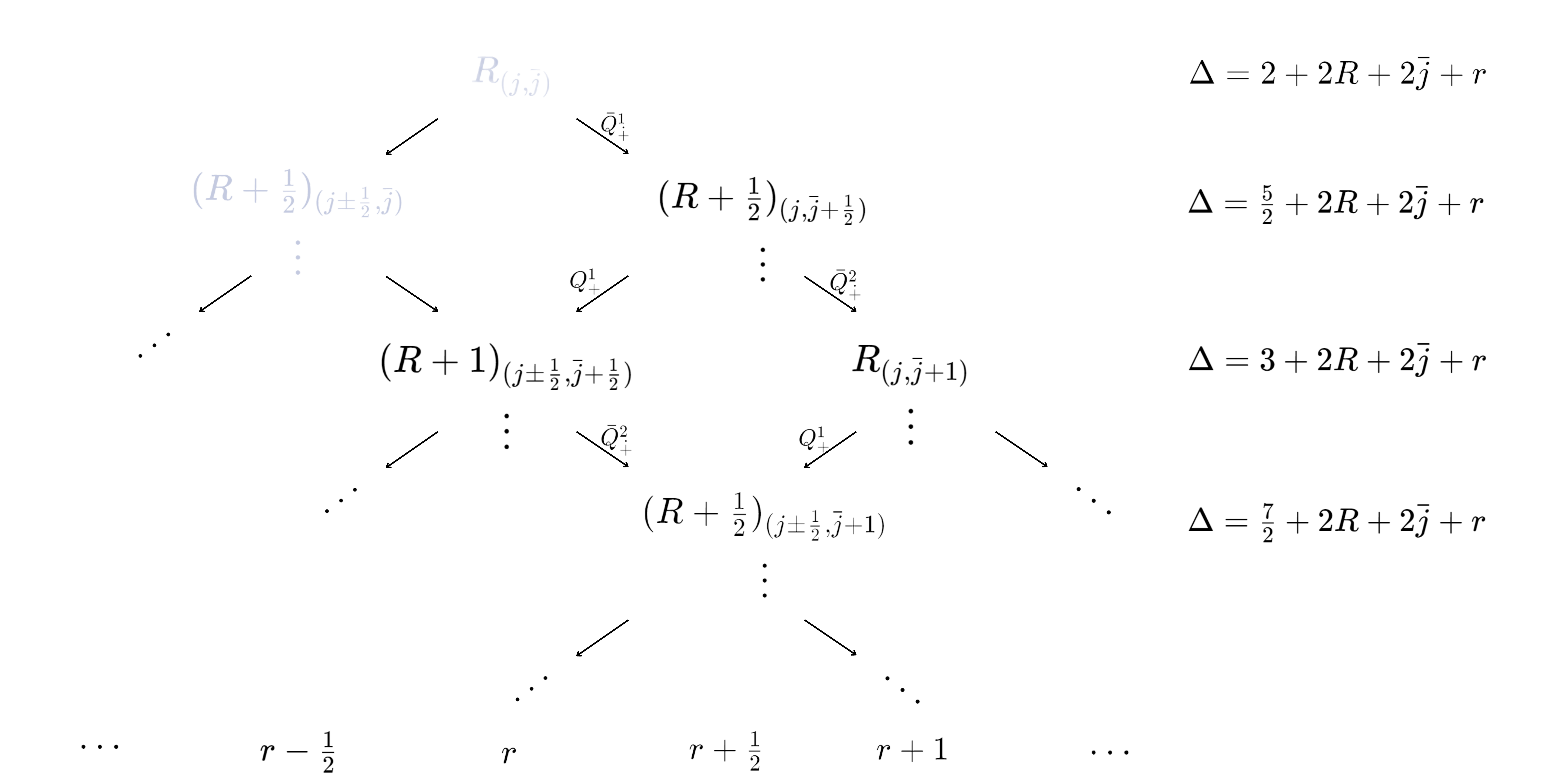}
\caption{A portion of a $\bar\CC_{R,r(j,\bar{j})}$ multiplet, with conformal primaries having $\delta=0$ states emphasized in black. As described in the main text, these states can be obtained by acting with $\bar Q^1_{\dot+}$ along with non-vanishing combinations of $Q^1_{\alpha}$ and $\bar Q^2_{\dot+}$. As we will explain in our counting discussion below, for our particular operators of interest, we can focus on supercharges with \lq\lq$+$" and \lq\lq$\dot+$" Lorentz weights. Hence, we have only included the action of $Q^1_+$, $\bar Q^1_{\dot+}$, and $\bar Q^2_{\dot+}$. Note that in the case of $\hat\CC_{R(j,\bar j)}$, some of the above states are absent due to additional shortening conditions. Of the emphasized states in the figure, there is an $(R+1)_{(j-{1\over2},\bar j+{1\over2})}$ and an $(R+{1\over2})_{(j-{1\over2},\bar j+1)}$ state missing for $\hat\CC$ relative to $\bar\CC$.}\label{Cbardiag}
\end{figure}

Therefore, we can rephrase our problem as one of finding operators satisfying $\delta=0$ built out of products of fields in the chiral part of the vector multiplet. From table \ref{FVDeltaQDelta}, we see that we can only use the fields $\phi$, $\lambda^1_{\alpha}$, and $\partial_{\alpha\dot+}$ (the anti-holomorphic fields cannot appear in $\bar\CD_{0(0,0)}^{\times N}$). Moreover, from table \ref{FVDeltaQDelta}, we can see that if such operators sit in $\hat\CC$ or $\bar\CC$ multiplets, they must be descendants involving the action of $\bar Q^1_{\dot+}$ on the primary (so as to obtain $\delta=0$) possibly accompanied by any non-vanishing combination of the $\delta=0$ supercharges, $Q^1_{\alpha}$, $\bar Q^1_{\dot-}$, and $\bar Q^2_{\dot+}$ (see figure \ref{Cbardiag} for a diagrammatic representation of the $\delta=0$ operators). Note that the $\bar\CC$ and $\hat\CC$ shortening conditions imply that acting with $\bar Q^1_{\dot-}$ (along with $\bar Q^1_{\dot+}$) does not yield a new operator, and we can discard it.\footnote{In the case of $\bar\CD$ and $\bar\CE$ multiplets, the primary and the various $Q^1_{\alpha}$ descendants have $\delta=0$.}

We define the above set of operators and their multiplets to constitute the \lq\lq quasi-chiral" sector of the free vector multiplet (note that in this definition, the chiral sector is a subset of the \lq\lq quasi-chiral" sector). These operators are the full set of short multiplets appearing in the $\bar\CD_{0(0,0)}^{\times N}$ OPEs.

We will restrict our analysis to the $s=-1$ part of the quasi-chiral sector, where
\begin{equation}\label{sPrimary}
s_{ \CO }:=R_{ \CO }+\bar j_{ \CO }-j_{ \CO }=-1~,
\end{equation} 
for the highest $SU(2)_R$ and Lorentz-weight primary, $ \CO$. Such multiplets are of the form
\begin{equation}\label{sIR}
\bar\CC_{{p\over2},n+{p\over2}(k,k-1-{p\over2})}~,\ \ \ \hat\CC_{{p\over2}(k,k-1-{p\over2})}~,\ \ \ n\in\mathbb{Z}_{>1}~,\ p\in\mathbb{Z}_{\ge0}~.
\end{equation}
These degrees of freedom include all the complex higher-spin multiplets (the $\hat\CC$ multiplets in \eqref{sIR} with $p=0$), but also include infinitely many more. As we will see, these multiplets take part in a particularly simple complex geometrical construction that extends the Coulomb branch to an infinite-dimensional space.

\begin{table}[]
\centering
\begin{tabular}{|c|ccccc|cccccccccc|}
\hline
$s=-1$ &
  $\bar{Q}^{2}_{\dot{-}}$ &
  $Q^{2}_{+}$ &
   &
   &
   &
  $\lambda^{2}_{+}$ &
  $F_{+ +}$ &
  $\partial_{+ \dot{-}}$ &
  $\bar{\lambda}^{2}_{\dot{-}}$ &
  $\bar{F}_{\dot{-} \dot{-}}$ &
   &
   &
   &
   &
   \\
$s=0$ &
  $\bar{Q}^{2}_{\dot{+}}$ &
  $Q^{2}_{-}$ &
  $\bar{Q}^{1}_{\dot{-}}$ &
  $Q^{1}_{+}$ &
   &
  $\lambda^{2}_{-}$ &
  $\lambda^{1}_{+}$ &
  $F_{- +}$ &
  $\phi$ &
  $\partial_{+  \dot{+}}$ &
  $\partial_{- \dot{-}}$ &
  $\bar{\phi}$ &
  $\bar{\lambda}^{2}_{\dot{+}}$ &
  $\bar{\lambda}^{1}_{\dot{-}}$ &
  $\bar{F}_{\dot{-} \dot{+}}$ \\
$s=1$ &
  $\bar{Q}^{1}_{\dot{+}}$ &
  $Q^{1}_{-}$ &
   &
   &
   &
  $\lambda^{1}_{-}$ &
  $F_{- - }$ &
  $\partial_{- \dot{+}}$ &
  $\bar{\lambda}^{1}_{\dot{+}}$ &
  $\bar{F}_{\dot{+} \dot{+}}$ &
   &
   &
   &
   &
   \\ \hline
\end{tabular}
\caption{$s$ quantum numbers of the $\CN = 2$ Poincar\'e supercharges and fields in the free vector multiplet, $\bar\CD_{0(0,0)}\oplus\CD_{0(0,0)}$.}
\label{FVsQs}
\end{table}

At the level of the $\delta=0$ descendant operators of the multiplets in \eqref{sIR}, we must have $\delta=s=0$ (this statement follows from the supercharge quantum numbers in tables \ref{FVDeltaQDelta} and \ref{FVsQs}). As a result, all operators in the multiplets of \eqref{sIR}  that contribute to the index must be built from
\begin{equation}\label{IRdof}
\phi~,\ \ \ \lambda^1_+\sim Q^1_+\phi~,\ \ \ \partial_{+\dot+}~.
\end{equation}
In the IR, this sector is quite closely related to the Schur sector. The difference is that we swap $\bar\lambda^1_{\dot+}$ for $\phi$. Since the MAD theory is essentially a \lq\lq Coulombic" theory, this maneuver yields considerable information.

This logic also shows that the $\bar\CC$ and $\hat\CC$ multiplets with $s=-1$ in \eqref{sIR} arising from the quasi-chiral sector are exhaustive. The reason is that, to find such a multiplet outside the quasi-chiral sector, we would need to use fields in the anti-chiral $\CD_{0(0,0)}$ multiplet having $\delta=s=0$. However, a quick glance at tables \ref{FVDeltaQDelta} and \ref{FVsQs} reveals there are no such fields.

\bigskip
\noindent
{\bf Upshot:} At generic points on the Coulomb branch of the MAD theory, the full set of multiplets with quantum numbers in \eqref{sIR} appear in the $n+p+1$-fold OPE, $\bar\CD_{0(0,0)}^{\times(n+p+1)}$, i.e.
\begin{equation}\label{UpshotOPE}
\bar\CD_{0(0,0)}^{\times(n+p+1)}\ni\bar\CC_{{p\over2},n+{p\over2}(k,k-1-{p\over2})}~,\ \ \ \hat\CC_{{p\over2}(k,k-1-{p\over2})}~,\ \ \ n\in\mathbb{Z}_{>1}~,\ p\in\mathbb{Z}_{\ge0}~.
\end{equation}

\subsec{Counting the multiplicities of quasi-chiral multiplets  }
Let us now proceed to find the spectrum of multiplets in \eqref{sIR}. To that end, let us focus on their level-one $\bar Q^1_{\dot+}$ $\delta=0$ descendants. The above discussion implies that such an operator, $X$, takes the general form
 \be\label{Xform}
 X = \sum_{\bm\ell,\bm a} x_{\bm\ell,\bm a} \partial_{+\dot+}^{\ell_1}\lambda^1_+ 
 \partial_{+\dot+}^{\ell_2}\lambda^1_+ \cdots \partial_{+\dot+}^{\ell_{p+1}}\lambda^1_+ 
 \partial^{a_1}_{+\dot+}\phi \partial^{a_2}_{+\dot+}\phi \cdots \partial^{a_n}_{+\dot+}\phi
 ~,
\ee
with coefficients $x_{\bm\ell,\bm a}\in \mathbb C$ and quantum numbers
\begin{eqnarray}
&R_X =\frac{p+1}{2}~,  \qquad\qquad\qquad\qquad\qquad\qquad &   r_X = \frac{p+1}{2}+n~,
\cr
& j_X =k=\frac12\Big( p+1+\sum\ell_j+\sum a_i\Big)~, \quad & 
\bar j  _X=\frac12\Big(   \sum\ell_j+\sum a_i\Big)=k  -\frac{p +1 }{2}~  .\qquad
\end{eqnarray}

To count the number of corresponding multiplets, we need to ensure that $X$ is a conformal primary and that it is not a $Q^1_+$ or $\bar Q^2_{\dot+}$ descendant. In particular
\begin{equation}\label{constraints}
K_{\mu}X(0)=0~,\ \ \ X\ne Q^1_+Y_{p,-1}~, \ \ \ X\ne \bar Q^2_{\dot+}X_{p,2}~.
\end{equation}
The last two conditions imply that $X$ is the $\delta=0$ state at level-one. Note that we need not impose conditions involving the other $\delta=0$ supercharge, $Q^1_-$, because its action cannot generate an operator of the form \eqref{Xform} with all $\dot+$ indices.

\subsubsec{The $p=0$ case}
Let us begin by studying the $X$ operators for $p=0$. The general expression in \eqref{Xform} becomes
\be\label{Xform1}
X = \sum_{\ell,\bm a} x_{\ell,{\bm{a}}} \partial_{+\dot+}^{\ell}\lambda^1_+ \partial^{a_1}_{+\dot+}\phi \partial^{a_2}_{+\dot+}\phi\cdots \partial^{a_n}_{+\dot+}\phi~,\qquad 
x_{\ell,{\bm{a}}}\in \mathbb C~,
\ee
with quantum numbers
\be
R=\frac12~, \qquad j=k=\frac12\Big( 1+\ell+\sum a_i\Big)~, \qquad \bar j =k-\frac12~, \qquad r=\frac12+n~.
\ee
In this case, the conditions in \eqref{constraints} simplify
\begin{equation}\label{constraints0}
K_{\mu}X(0)=0~,\ \ \ X\ne \bar Q^2_{\dot+}X_2~,
\end{equation}
where we have defined $X_2:=X_{0,2}$. Indeed, we can drop the constraint $X\ne Q^1_+Y_{-1}$ (where $Y_{-1}:=Y_{0,-1}$) since $X$ has $R=1/2$. On the other hand, in a $\bar\CC$ or $\hat\CC$ multiplet, a $\delta=0$ conformal primary of the form $Q^1_+Y_{-1}$ necessarily has $Y_{-1}=\bar Q^1_{\dot+} \CO $, where $ \CO $ is the highest $SU(2)_R$ and Lorentz weight primary (see figure \ref{Cbardiag}).\footnote{One may also consider the $\delta=0$ state $Q^1_+\bar Q^2_{\dot+}\bar Q^1_{\dot+} \CO $. However, this state differs from one of the form $\bar Q^2_{\dot+}Q^1_+\bar Q^1_{\dot+} \CO $ by a total derivative and therefore may be neglected in the counting of conformal primaries (see figure \ref{Cbardiag}).} Therefore, $Q^1_+Y_{-1}$ has $R\ge1$, which is a contradiction.

Neglecting for a moment the constraints in \eqref{constraints0}, the number of $X$ operators in \eqref{Xform1} is clearly
 \be
 N_X(n,k)=\sum_{\ell=0}^{2k-1}S(2k-1-\ell,n)~.
 \ee
Here we define $S(2k-1-\ell,n)$ to be the number of ways to partition $2k-1-\ell$ into $n$ non-negative integers.  For more details on the corresponding generating function, see Appendix \ref{PartInt}. 

To implement the $\bar Q^2_{\dot+}$ constraint in \eqref{constraints0}, we need to subtract operators of the form $\bar Q^2_{\dot+}X_2$ with
\be\label{X2space}
X_2 = \sum_{\bm\ell, \bm a} x_{2,\bm\ell,{\bm a}}\partial_{+\dot+}^{\ell_1 }\lambda^1_+\partial_{+\dot+}^{\ell_2 }\lambda^1_+ \partial^{a_1}_{+\dot+}\phi \partial^{a_2}_{+\dot+}\phi \cdots \partial^{a_{n-1}}_{+\dot+}\phi~, \qquad \sum a_j+\sum \ell_j =2k-2~.
\ee
From the space of operators in \eqref{X2space}, we will also need to subtract operators satisfying $\bar Q^2_{\dot+}X_2=0$. As we will show in section \ref{cohomology}, the $\bar Q^2_{\dot+}$ cohomology on the space of $\delta=0$ operators in $\bar\CC$ and $\hat\CC$ multiplets is trivial. As a result, we must subtract operators of the form $X_2=\bar Q^2_{\dot+}X_3$ for
\be
X_3 = \sum_{\bm \ell, \bm a} x_{3,{\bm\ell,\bm a}}\partial_{+\dot+}^{\ell_1 }\lambda^1_+\partial_{+\dot+}^{\ell_2 }\lambda^1_+ \partial_{+\dot+}^{\ell_3 }\lambda^1_+ \partial^{a_1}_{+\dot+}\phi \partial^{a_2}_{+\dot+}\phi \cdots \partial^{a_{n-2}}_{+\dot+}\phi~, \qquad \sum a_j +\sum \ell_j=2k-3~.
\ee 
Again, from the space of $X_3$ operators, we should subtract $\bar Q^2_{\dot+}$-closed operators. By the discussion in section \ref{cohomology}, the corresponding cohomology turns out to be trivial and so we must subtract operators of the form $X_3=\bar Q^2_{\dot+}X_4$ and so on. Defining
\be\label{Xrcons1}
X_{ q } = \sum_{\bm \ell, \bm a}x_{ q ,\bm\ell,\bm a} \partial_{+\dot+}^{\ell_1 }\lambda^1_+ \cdots \partial_{+\dot+}^{\ell_{ q } }\lambda^1_+ \partial^{a_1}_{+\dot+}\phi \partial^{a_2}_{+\dot+}\phi \cdots \partial^{a_{n- q +1}}_{+\dot+}\phi~, \qquad \sum a_j +\sum \ell_j=2k- q ~,
\ee 
and denoting the set of such operators as $\Xi_ q $,
we find the following exact sequence (with $q_{\rm max}$ a finite upper bound we will specify in more detail below)
\be\label{Grseq}
0\xrightarrow{   \bar Q^2_{\dot+}} \Xi_{q_{\rm max}} \xrightarrow{   \bar Q^2_{\dot+}} \Xi_{q_{\rm max}-1}   \xrightarrow{   \bar Q^2_{\dot+}}\cdots   \xrightarrow{   \bar Q^2_{\dot+}} \Xi_2 \xrightarrow{   \bar Q^2_{\dot+}}
 \Xi_1  \xrightarrow{   \bar Q^2_{\dot+}}  \Xi_0  \xrightarrow{   \bar Q^2_{\dot+}} 0~, \ \ \ X_i\in\Xi_i~,
\ee
Note that $X\equiv X_1\in\Xi_1$, and let us define $N_{X_i}:={\rm dim}_{\mathbb{C}}(\Xi_i)$ with $N_{X_1}=N_X$. Then, we have
\beqn
 N_{X_{ q }}(n,k)  &=&
\sum_{\ell_1=0}^{2k-2}\sum_{\ell_2=0}^{\ell_1-1} \cdots\sum_{\ell_{ q }=0}^{\ell_{ q -1}-1} 
S(2k- q -\ell_1-\ell_2-\cdots -\ell_{ q },n- q +1)~.
\eeqn

From this discussion, we see that, after imposing the constraint $X\ne\bar Q^2_{\dot+}X_2$, the number of $X$ operators is
\begin{equation}\label{NXcons1}
N(p=0,n,k)=\sum_{ q =1}^{ q _{\rm max}}(-1)^{ q +1}N_{X_{ q }}(n,k)= \sum_{ q =1}^{ q _{\rm max}}  \sum_{\ell =0}^{2k- q } (-1)^{ q +1} R(\ell, q )S(2k- q -\ell ,n- q +1) ~,
\end{equation}
where $R(\ell, q )$ is the number of partitions of $\ell$ into $ q $ {\it distinct} non-negative integers. See appendix \ref{PartInt} for more details. Here we  have defined $ q _{\rm max}$ to be the maximal value of $ q $ such that $R(2k- q , q )\ne0$.

Finally, we should impose the condition that $X$ is a primary. Doing so we find that the total number of such primaries, and therefore the total number of $\bar\CC_{0,n(k,k-1)}$ and $\hat\CC_{0(k,k-1)}$ multiplets, is
\begin{eqnarray}
N_{\bar\CC_{0,n(k,k-1)}}&=&N(0,n,k)-N(0,n,k-1/2)~, \ \ \ n\in\mathbb{Z}_{>1}~,\cr N_{\hat\CC_{0,(k,k-1)}}&=&N(0,1,k)-N(0,1,k-1/2)=\begin{cases}
1~,\ k\in\mathbb{Z}_{>0}~,\\ 
0~, \ \text{otherwise}~ ,
\end{cases}~
\end{eqnarray}
where we have used the fact that conformal descendants are total derivatives. 
 
In the next section, we generalize the above results to the case of $p>0$ and subsequently prove that the $\bar Q^2_{\dot+}$ cohomology discussed above is indeed trivial.

\subsubsec{The general case}
Let us now consider the general case with $p\ge0$. For $p>0$, we must also impose the $X\ne Q^1_+Y_{p,-1}$ constraint in \eqref{constraints}. Indeed, the logic below \eqref{constraints0} no longer holds since now $X$ has $R\ge1$. 

Let us define the following generalization of \eqref{Xrcons1}
\be\label{Xrgen}
X_{p, q } = \sum_{\bm \ell, \bm a}x_{  p,q,\bm\ell,\bm a} \partial_{+\dot+}^{\ell_1 }\lambda^1_+ \cdots \partial_{+\dot+}^{\ell_{ q +p} }\lambda^1_+ \partial^{a_1}_{+\dot+}\phi \partial^{a_2}_{+\dot+}\phi \cdots \partial^{a_{n- q +1}}_{+\dot+}\phi~, \ \ \  \sum a_j +\sum \ell_j=2k- q -p~,
\ee 
and denote the space of such operators as $\Xi_ {p,q} $. Note that $X_{ q }$ in \eqref{Xrcons1} is now $X_{0, q }$. Imposing the constraint $X\ne\bar Q^2_{\dot+}X_{p,2}$ and counting solutions as before leads to the following generalization of \eqref{NXcons1}
\begin{equation}\label{NXgen}
N(p,n,k)=\sum_{ q =1}^{ q _{\rm max}}  \sum_{\ell =0}^{2k- q -p} (-1)^{ q +1} R(\ell, q +p)S(2k- q -p-\ell ,n- q +1)~,
\end{equation}
where $ q _{\rm max}$ is defined as the largest $ q $ such that $R(2k- q -p, q +p)\ne0$. This expression counts the number of operators of interest prior to imposing the conformal primary constraint and prior to imposing the $X\ne Q^1_+Y_{p,-1}$ constraint. As in the $p=0$ case discussed in the previous subsection, we again assume the triviality of the $\bar Q^2_{\dot+}$ cohomology on the space of $\delta=0$ operators in $\bar\CC$ and $\hat\CC$ multiplets. As indicated above, this assumption will be justified in section \ref{cohomology}.

To understand the imposition of the remaining constraints, let us consider $p=1$. In this case, we have to subtract operators of the form $Q^1_+Y_{1,-1}$, where $Y_{1,-1}=\bar Q^1_{\dot+} \CO $. Here $ \CO $ has quantum numbers $R=0$, $j=k-1/2$, $\bar j=k-3/2$, and $r=n+1$. These are precisely the $p=0$ operators we encountered in the previous subsection. As a result, the number of $X$ operators after imposing $X\ne\bar Q^2_{\dot+}X_3$ and $X\ne Q^1_+Y_{1,-1}$ is
\begin{equation}\label{subtractp0}
N'(1,n,k)=N(1,n,k)-N(0,n+1,k-1/2)~.
\end{equation}

For any $p>0$, we easily find the following generalization of \eqref{subtractp0} (assuming the triviality of the $Q^1_+$ cohomology on the space of $\bar\CC$ and $\hat\CC$ multiplets)
\begin{equation}\label{subtractpGen}
N'(p,n,k)=\sum_{i=0}^{p}(-1)^iN(p-i,n+i,k-i/2)~.
\end{equation}
Finally, imposing the condition of being a conformal primary yields the following spectrum
\begin{eqnarray}\label{IRspectrum}
N_{\bar\CC_{{p\over2},n+{p\over2}(k,k-1-{p\over2})}}&=&N'(p,n,k)-N'(p,n,k-1/2)~, \ \ \ n\in\mathbb{Z}_{>1}~,\cr N_{\hat\CC_{{p\over2},(k,k-1-{p\over2})}}&=&N'(p,1,k)-N'(p,1,k-1/2)~.
\end{eqnarray}
The same results can also be obtained from a double complex that we describe in Appendix~\ref{AppB}.

One feature of this spectrum is that, for fixed $R$, $j$, and $\bar j$, the number of multiplets eventually becomes constant as we increase the $U(1)_r$ charge (we say that the spectrum \lq\lq stabilizes"). Said differently, beyond a particular value of the $U(1)_r$ charge (i.e., for $r\ge r_{k,p}=n_{k,p}+{p\over2}$) the SCPs of the $\bar\CC$ multiplets in \eqref{IRspectrum} factorize as follows
\be\label{FVfactorize}
\CO_{r_{k,p}+m}=\CO_{r_{k,p}} \phi^{m}~, \qquad
\CO_r \in \bar\CC_{{p\over2},r(k,k-1-{p\over2})}~,  \quad m \in \mathbb N~.
\ee
In other words, the spectrum factorizes in the large-$ r$ limit.

To explicitly see the above stabilization and factorization of the spectrum, we note that for $b\ge a$, $S(a,b)=P(a)$, where $P(a)$ is the number of ways to partition the integer $a$ into arbitrary positive integers. Therefore, taking $r\ge r_{k,p}=2k-1-p/2$ or, equivalently, $n \ge n_{k,p}= 2k-1-p$, \eqref{NXgen} becomes
\be\label{NXgen2}
N(p,n\ge n_{k,p},k)=\sum_{ q =1}^{ q _{\rm max}}  \sum_{\ell =0}^{2k- q -p} (-1)^{ q +1} R(\ell, q +p)P(2k- q -p-\ell)~.
\ee 
In particular, we see that $N(p,n\ge n_{k,p},k)$ is independent of $n$. Moreover, $N'(p,n\ge n_{k,p},k)$ and $N'(p,n\ge n_{k,p},k-1/2)$ are also independent of $n$. Therefore,
\begin{equation}
N_{\bar\CC_{{p\over2},n+{p\over2}(k,k-1-{p\over2})}}=N_{\bar\CC_{{p\over2},n_{k,p}+{p\over2}(k,k-1-{p\over2})}}~, \ \ \ \forall\ n\ge n_{k,p}=2k-1-p~.
\end{equation}

Note that in the particular case of $p=0$, the formulae can be further simplified 
\be
N(p=0,n\ge 2k-1,k)= \sum_{ q =1}^{ q _{\rm max}}  \sum_{\ell =0}^{2k- q } (-1)^{ q +1} R(\ell, q )P(2k- q -\ell  ) 
=P(2k)-\delta_{k,0}~,
\ee
 where we use the identity \eqref{idPart}.  As a result, 
 \be
 N_{\bar\CC_{0,n(k,k-1)}}=P(2k)-P(2k-1)~, \qquad n\ge 2k-1~, \quad k\ge 1~.
 \ee

More generally, we expect a similar stabilization and factorization discussed above to hold for all $\bar\CC_{R,r(j,\bar j)}$ multiplets in the free vector theory for fixed $R$, $j$, and $\bar j$ and sufficiently large $r$ (although the particular $U(1)_r$ threshold for different classes of multiplets will be different). The reason is that fixing $R$ does not allow us to increase the number of gauginos in our $\delta=0$ operators (recall from table \ref{FVDeltaQDelta} that all $\lambda^2_{\alpha}$ and $\bar\lambda^2_{\dot\alpha}$ gauginos have $\delta>0$). Therefore, to increase $r$, we must increase the number of $\phi$'s. Moreover, from table \ref{FVDeltaQDelta} we also see that any right-handed spinor index contractions increase $\delta$. As a result, we cannot increase the number of derivatives or $\bar F_{\dot+\dot+}$ operators for fixed spin, and we are led to a factorization of the form \eqref{FVfactorize} if we multiply by sufficiently many $\phi$.

Let us conclude by reminding the reader that our construction of the spectrum in \eqref{IRspectrum} depended on the triviality of the $Q^1_+$ and $\bar Q^2_{\dot+}$ cohomologies. In the next subsection we give a proof of this statement along with a geometrical construction of the multiplets in \eqref{IRspectrum}.

\subsection{Triviality of the $\bar Q^2_{\dot+}$ and $Q^1_+$ cohomologies  }\label{cohomology}
In our arguments leading up to \eqref{IRspectrum}, we assumed that the $\bar Q^2_{\dot+}$ and $Q^1_+$ cohomologies were trivial when acting on the space of $\delta=0$ operators in $\bar\CC$ and $\hat\CC$ multiplets. In particular, we have the following exact sequence
\begin{eqnarray}
0&\xrightarrow{   \bar Q^2_{\dot+}}& \Xi_{p,q_{\rm max}} \xrightarrow{   \bar Q^2_{\dot+}} \Xi_{p,q_{\rm max}-1}   \xrightarrow{   \bar Q^2_{\dot+}}\cdots   \xrightarrow{   \bar Q^2_{\dot+}} \Xi_{p,2} \xrightarrow{   \bar Q^2_{\dot+}}
 \Xi_{p,1}  \xrightarrow{   \bar Q^2_{\dot+}}  \Xi_{p,0}  \xrightarrow{   \bar Q^2_{\dot+}} 0~, \ \ \ X_{p,i}\in\Xi_{p,i}~,\ \ \ \ \ \ \ \ \ 
\end{eqnarray}
where the level-one $\bar Q^1_{\dot+}$ descendant of a multiplet in \eqref{IRspectrum} satisfies $X\in\Xi_{p,1}$. Note that there is a similar exact sequence for $Q^1_+$ (see figure \eqref{doublecomplex} in Appendix \ref{AppB} for further details).

Since the chiral field $\phi$ forms a ring (i.e., its product with itself has no singularities), one can formally think of it as corresponding to a complex variable, $\phi\sim z$ (recall that the ring of chiral operators is isomorphic to the ring of holomorphic polynomials on the Coulomb branch). More generally, the products of any combinations of $\partial^i_{+\dot+}\phi$ and $\partial^j_{+\dot+}\lambda^1_+$ are non-singular (although potentially vanishing if sufficiently many fermions are involved) and so we can try to give them a simple geometrical interpretation. To that end, let us make the formal identifications
\be\label{geoId1}
z^i\leftrightarrow\p^i_{+\dot+} \phi~, \qquad dz^i \leftrightarrow\p^i_{+\dot+}\lambda^1_+~, \qquad \  d=dz^i \wedge \p_i\leftrightarrow Q_{ +}^1~,\qquad \partial_i:=\partial_{z^i}~,
\ee   
where $z^i$ is the $i$-th complex variable (not the $i$-th power of $z$). In particular, the $z$ we introduced above \eqref{geoId1} is now $z^0$. It is then easy to check that \eqref{geoId1} is indeed compatible with the action of supersymmetry\footnote{In this subsection, to simplify the equations, we have rescaled the fields and supercharges appropriately in order to remove any overall numerical factors.}
\be
dz^i \leftrightarrow Q_{ +}^1\p^i_{+\dot+} \phi =\p^i \lambda^1_+~, \qquad 0=ddz^i \leftrightarrow Q_{ +}^1 \p^i_{+\dot+}\lambda^1_+=0~.
\ee

This construction also allows us to write down multi-field operators as differential forms
\be\label{Xomega}
X \leftrightarrow \omega=\sum_{\bm k} P_{\bm k} (z^i) dz^{k_1}\wedge dz^{k_2} \cdots~,
\ee
where $P_k$ is a polynomial in the $z^i$. So if $Q_+^1X=0$, namely $d\omega=0$, then we can define\footnote{Here we assume that $\omega$ is at least a 1-form, otherwise the integration is meaningless. If $\omega$ is a 0-form/function, then $d\omega=0$ means that $\omega$ is a constant, corresponding to the identity operator $X\sim 1$. } 
\be
\Omega(z)=\int_0^z \omega=\int_0^z \sum_{\bm k} P_{\bm k}(z^i) dz^{k_1}\wedge dz^{k_2} \cdots~, \qquad
\omega=d\Omega~.
\ee
Note that closedness, $d\omega=0$, ensures the path-independence of the integral, while the absence of a singularity in $P_{\bm k}$ ensures that $\omega$ is integrable. Therefore, the fact  that $Q_{ +}^1$ cohomology is trivial is just the geometric fact that the (complexification of the) $\mathbb{R}^n$ de Rham cohomology is trivial.\footnote{More precisely, this is the Dolbeault cohomology $H^{\bullet,0}(\mathbb C^n)$.}
 
Next let us consider the action of $\bar Q_{\dot+}^2$. We claim we can associate it with the interior product
\be
\iota_{v} \leftrightarrow \bar Q_{\dot+}^2~, \qquad v=\sum_i z^{i+1}\p_i~, \qquad
\iota_{v}\iota_{v}=0~.
\ee
Indeed, this operator is consistent with the action of the supercharge
 \be\label{geoId2}
 \iota_{v} dz^i=z^{i+1} \leftrightarrow \bar Q_{\dot+}^2 \p^i_{+\dot+} \lambda^1_+= \p^{i+1}_{+\dot+}\phi, \qquad
  \iota_{v}  z^i=0 \leftrightarrow
\bar Q_{\dot+}^2 \p^i_{+\dot+} \phi=0~.
 \ee 
Finally, we can identify $\partial_{+\dot+}$ with the Lie derivative with respect to $v$
\be
\mathcal L_v dz^j=\p_i  v^j  dz^i=dz^{j+1}~, \qquad
\mathcal L_v  z^j=z^{i+1}\p_i z^j=z^{j+1}~.
\ee
Moreover, the above equations are consistent with the supersymmetry algebra
\be
\{\bar Q_{\dot+}^2 , Q_+^1\}=\p_{+\dot +} \leftrightarrow
d\iota_{v} +\iota_{v}  d=\mathcal L_v~,
\ee
where we have used the Cartan identity.

Now we would like to show that $\bar Q_{\dot+}^2$ has trivial cohomology when acting on the forms defined in \eqref{Xomega}.\footnote{Except for the case of 0-forms depending only on $z^0$. This situation corresponds to chiral operators, $X=\phi^n$ ($n$ can be zero, in which case $X$ is the identity operator), where $\bar Q_{\dot+}^2X=0$, but $X$ cannot be written as $X=\bar Q_{\dot+}^2 X'$.} In other words, we would like to show that $\iota_{v} \omega=0$ implies $\omega{=}\iota_{v}  \Omega$. We can take a pointwise approach to the problem.

To that end, first consider a point where $ v\neq 0$. Using the formula $\iota_v \star\alpha =(-1)^{  |\alpha|}\star( V \wedge  \alpha)$ where $V$ is a 1-form associated with $v$, the question can be rephrased as one of showing
\be 
\iota_v \omega = \pm \star( V \wedge \star\omega)=0 \quad \Rightarrow\quad
\star\omega= \pm( V \wedge \star \Omega)~,
\ee
where the sign is not important. Note that $\star \omega=0$ iff $\omega=0$. As a result, the question can be further rephrased as asking if we can show whether $V\wedge U=0$ implies $U=V\wedge W$. Locally and point-wise, we can always  choose an orthogonal  frame basis, $e_i$, such that  $e_1 =V$,  and  write 
\be
U=V\wedge W+Y~,
\ee
where $Y$ only has components  $e_2$, $e_3$, $\cdots$. In other words, $Y=\sum_i Y_i e_{i_1} \wedge e_{i_2} \cdots $, with indices $1<  {i_1}<  {i_2}< \cdots $. Then $V \wedge U=V \wedge Y = \sum_i Y_i e_1\wedge e_{i_1} \wedge e_{i_2} \cdots$, which is zero iff $Y_i=0$. Therefore, we have shown that $V \wedge U=0$ iff $U=V \wedge W$. So, if $v\ne0$, $\iota_{v} \omega=0$ implies $\omega\stackrel{ }{=}\iota_{v}  \Omega$. 

Next consider a point where $v=0$ and thus $\iota_v \omega=0$ holds identically  for all $\omega$.  In this case, we can consider a neighbourhood of $v=0$ and approach $v=0$ by taking limits. The vector field $v$ vanishes at points with $z^1=z^2 =\cdots =0$. We can check our statement away from these points; if it holds there, we expect our statement holds at the zeros as well. However, the argument fails if $\iota_v \omega$ is independent of $z^1$, $z^2$,  $\cdots$. This situation occurs if $\omega$ is independent of $z^1$, $z^2$, $\cdots$ and $dz^0$, $dz^1$, $dz^2$, $\cdots$ (recall that $\iota_v dz^i=z^{i+1}$).  In other words, we must be in the situation where $\omega=f(z^0)$ is  just a function of $z^0$.\footnote{Note that if  a 0-form / function $\omega$ depends on $z^1$, $z^2$, $\cdots$, then we can always write $\omega =z^i f(z)$  and $\iota_v \omega=0$ where $i\neq 0$. However, we have $\omega =\iota_v(dz^{i-1} f(z))$.} In this case, $\iota_v \omega=0$ holds identically, but $\omega=f(z^0)\neq\iota_v \Omega$, as the latter must depend on $z^1$, $z^2$, $\cdots$ (recall again that $\iota_v dz^i=z^{i+1}$). Physically, this situation corresponds to the case $X\sim \phi^n$ discussed above. In this case, $\bar Q_{\dot+}^2X=0$, but $X$ cannot be written as $ \bar Q_{\dot+}^2 X'$.
 
Therefore, we have shown that the $\bar Q_{\dot+}^2$-cohomology is indeed trivial when acting on the space of $\delta=0$ operators in $\bar\CC$ and $\hat\CC$ multiplets.\footnote{We have also checked this statement numerically by constructing the set of operators explicitly.} As a result, we have justified the computations leading up to \eqref{IRspectrum}.

We can also use the above language to give a geometrical construction of the $X\in\bar\CC_{{p\over2},n+{p\over2}(k,k-1-{p\over2})}$, $\hat\CC_{{p\over2}(k,k-1-{p\over2})}$ operators in \eqref{Xform} satisfying \eqref{constraints} (recall that $X$ is the $\bar Q^1_{\dot+}$ descendant of the $\CN=2$ superconformal primary). To that end, let $\omega$ be the form corresponding (via \eqref{Xomega}) to some $\delta=0$ operator with the quantum numbers of $X$. Define the equivalence class
\begin{equation}
[\omega]:=\left\{\omega'|\ \exists   \beta ~,\ {\rm s.t.}\ \omega'=\omega +  \CL_v \beta\right\}~,
\end{equation}
where $\beta$ is a polynomial form. Then, the multiplets of interest are in one-to-one correspondence with equivalence classes of $(p+1)$-forms, $[\omega]$, satisfying
\begin{equation}
0\not\in[\omega]~, \ \ \ d\omega'\neq 0~,\ \iota_v\omega'\ne0~,\ \ \ \forall\omega'\in[\omega]~.
\end{equation}
Alternatively, we can work with the level-three descendant $Q^1_+\bar Q^2_{\dot+}\bar Q^1_{\dot+}$ of the SCP (or any other $\delta=0$ operator), $Z\in\bar\CC_{{p\over2},n+{p\over2}(k,k-1-{p\over2})}$, $\hat\CC_{{p\over2}(k,k-1-{p\over2})}$. In this case, we have that $Z$ corresponds to an equivalence class of $(p+1)$-forms satisfying
\begin{equation}
0\not\in[\omega]~,\ \ \ \hat\omega,\tilde\omega\in[\omega]~, \ \ \ d\hat\omega=0~,\ \ \ \iota_v\tilde\omega=0~,\ \ \ \hat\omega\ne\tilde\omega~.
\end{equation}

In the next section we will reinterpret our IR discussion in the preceding sections in terms of UV degrees of freedom in the MAD theory. We then conclude with a highly non-trivial check of our results using the superconformal index.

 \newsec{The MAD theory: UV avatars}
In this section, we will connect the discussion of the deep IR in the previous section with the MAD theory. In particular, we will find UV avatars for the $\bar\CC$ and $\hat\CC$ multiplets studied above. Using the resulting MAD / free vector map, we then propose  a formula counting the full set of $\bar \CC$ multiplets with $s=-1$ in the MAD SCFT. We conclude by checking that the formula is consistent with the superconformal index. 
\subsec {The MAD / free vector map}\label{MADfvmap}
In this subsection, we would like to explain the UV origins of the IR spectrum found in \eqref{IRspectrum}. To that end, recall that the MAD theory has a dimension $6/5$ $\CN=2$ chiral primary, $\sO_{6/5}$. This operator flows to $\phi$ under the RG flow to the Coulomb branch\footnote{Other operators, like $\sO_{6/5}^2$, also flow to $\phi$ at leading order in the IR (this is no surprise since $U(1)_r$ is spontaneously broken). However, we can subtract off a term proportional $\sO_{6/5}$ in the UV to obtain an operator that flows to $\phi^2$ at leading order. In this sense, we will say $\sO_{6/5}^2$ flows to $\phi^2$.\label{mapclear}}
\begin{equation}\label{mapE}
\sO:=\sO_{6/5}\in\bar\CE_{6/5}\longrightarrow\phi~,
\end{equation}
where we have relabelled the chiral primary for simplicity. It is then natural to study the UV operators corresponding to \eqref{IRdof}
\begin{equation}\label{MADcont}
\sO\longrightarrow\phi~,\ \ \ \sO_{+}:=\sO^1_{+}\sim Q^1_+\sO\longrightarrow\lambda^1_+~,\ \ \ \partial_{+\dot+}~,
\end{equation}
and to study the UV avatar of $X$ in \eqref{Xform} built from these degrees of freedom. Let us call this operator $X^{\text{MAD}}$.

More precisely, we would like to study $\bar Q^1_{\dot+}$ descendants, $X^{\text{MAD}}$, of UV $\bar\CC^{\text{MAD}}_{R,r(j,\bar j)}$ multiplets with primaries, $ \CO ^{\text{MAD}}$, satisfying
\begin{eqnarray}\label{COMADqn}
s_{ \CO ^{\text{MAD}}}&=&R_{ \CO ^{\text{MAD}}}+\bar j_{ \CO ^{\text{MAD}}}-j_{ \CO ^{\text{MAD}}}=-1~, \cr \delta_{ \CO ^{\text{MAD}}}&=&\Delta_{ \CO ^{\text{MAD}}}-\left(2R_{ \CO ^{\text{MAD}}}+2\bar j_{ \CO ^{\text{MAD}}}+r_{ \CO ^{\text{MAD}}}\right)=2~.
\end{eqnarray}
In terms of the $X$ quantum numbers, these constraints translate to
\begin{eqnarray}\label{XMADqn}
s_{X^{\text{MAD}}}&=&R_{X^{\text{MAD}}}+\bar j_{X^{\text{MAD}}}-j_{X^{\text{MAD}}}=0~,\cr \delta_{X^{\text{MAD}}}&=&\Delta_{X^{\text{MAD}}}-\left(2R_{X^{\text{MAD}}}+2\bar j_{X^{\text{MAD}}}+r_{X^{\text{MAD}}}\right)=0~.
\end{eqnarray}
The quantum numbers that make up $s$ are conserved along the RG flow ($SU(2)_R$ and Lorentz symmetry are preserved), and so any UV avatars of the operators in \eqref{IRspectrum} should satisfy these constraints. The second equations in \eqref{COMADqn} and \eqref{XMADqn} are the statements that the UV avatars of the IR multiplets are themselves $\bar\CC$ multiplets. This mapping is suggested by the general discussion in the introduction. In addition, in the next subsection, we will show that these statements are consistent with the superconformal index.

As a simple example of the construction we have in mind, let us consider the IR $\hat\CC_{0(1,0)}$ higher-spin multiplet  with primary described around \eqref{HSprim1}. Using the map in \eqref{mapE}, we expect a map of normal-ordered products\footnote{In the strongly coupled UV theory, a normal-ordered product is a spacetime-independent piece of an OPE.}
\begin{equation}\label{exmap}
\bar\CC^{\rm \, MAD}_{0,{7\over5}(1,0)}\ni\epsilon_{ij}\sO^i_{\alpha}\sO^j_{\beta}+\kappa'\sO\sO_{\alpha\beta} \ \longrightarrow\ \CO_{\alpha\beta}=\epsilon_{ij}\lambda^i_{\alpha}\lambda^j_{\beta}+\kappa\phi F_{\alpha\beta}\in\hat\CC^{\rm \, Free}_{0(1,0)}~,
\end{equation}
where $\sO^i_{\alpha}\sim Q^i_{\alpha}\sO$ and $\sO_{\alpha\beta}\sim\epsilon_{ij}Q^i_{\alpha}Q^j_{\beta}\sO$. Here we have chosen $\kappa'$ so that the LHS of \eqref{exmap} is an SCP. Note that $\kappa'\ne\kappa$, which is consistent with the fact that we expect conformal primaries and descendants to mix upon turning on a vev, $v=\langle\sO\rangle\ne0$, and flowing to the IR. Let us also note that, as in footnote \ref{mapclear}, the operator in \eqref{exmap} flows to a lower-dimensional operator than the one on the RHS at leading order in the IR. Indeed, substituting $v$ in \eqref{exmap}, we expect the leading IR operator to be proportional to $F_{\alpha\beta}$. However, we can remove this operator by modifying the RHS of \eqref{exmap} to include a linear combination with the level-two descendant $\sO_{\alpha\beta}$ of the $\bar\CE_{6/5}$ multiplet. Then, we expect \eqref{exmap} to hold (up to a shift by a descendant). It is in this sense that we think of the RHS of \eqref{exmap} as mapping to the LHS.\footnote{In this way, we can think of the higher-spin current on the RHS of \eqref{exmap} as having acquired anomalous dimension $2/5$ in the UV. Since $2/5\ll 3$, there is a sense in which the MAD theory has an \lq\lq approximate" higher spin symmetry.
Note that the higher spin $\hat\CC_{0(k,k-1)}$ analogs of \eqref{exmap} have anomalous dimension $2/5\ll 2k+1$ (since we only increase the number of derivatives on both sides of the map). For large $k$, this symmetry is in a sense \lq\lq parametrically conserved" (see \cite{Maldacena:2012sf} for a study of approximate higher-spin symmetries). It would be interesting to understand if this structure can be used to gain further insight into the MAD theory (and its higher-rank cousins).}

Similarly, at the level of the $X$ operator in \eqref{Xform1}, we expect the map of normal-ordered products
\begin{equation}\label{XmapUVIR}
X^{\text{MAD}}=\sum_{\ell,a} x'_{\ell,{a}} \partial_{+\dot+}^{\ell}\sO^1_+ \partial^{a}_{+\dot+}\sO \ \longrightarrow\sum_{\ell, a} x_{\ell,{a}} \partial_{+\dot+}^{\ell}\lambda^1_+ \partial^{a}_{+\dot+}\phi=X~,
\end{equation}
where we have used the $\delta=0$ degrees of freedom in \eqref{MADcont}. Related comments to those appearing below \eqref{exmap} apply to \eqref{XmapUVIR}.

In generalizing the above map, we should account for the fact that $\bar\CE_{6/5}$ satisfies fewer shortening conditions than the free $\bar\CD_{0(0,0)}$ multiplet. For example, $(Q^1)^2\sO\ne0$ has $\delta=0$ in the UV, but $(Q^1)^2\phi=0$ by equations of motion in the IR. One might imagine that such discrepancies could allow one to find a mismatch between UV and IR multiplets of the type we study.

However, we claim that only the operators in \eqref{MADcont}, which are in one-to-one correspondence with the contributing free vector operators (i.e., $\phi$, $\lambda_+$, and $\partial_{+\dot+}$), can be used in constructing the $X^{\text{MAD}}$ normal-ordered products. To understand this statement, first recall from tables \ref{FVDeltaQDelta} and \ref{FVsQs} that all derivatives have $s\ge0$ and $\delta\ge0$ except for $\partial_{+\dot-}$, which has $s=-1$ and $\delta=2$. Moreover, we see that only $\partial_{+\dot+}$ satisfies $s=\delta=0$. Next, note that conformal primaries in the $\bar\CE_{6/5}$ multiplet can only be obtained through the action of the $Q^i_{\alpha}$ supercharges. However, since $\delta_{Q^2_{\alpha}}=2$ and $\delta_{Q^1_{\alpha}}=0$, we can neglect $Q^2_{\alpha}$ and $\partial_{+\dot-}$. In addition,
\begin{equation}
s_{Q^1_+}=0~,\ \ \ s_{Q^1_{-}}=1~.
\end{equation}
As a result, all conformal primaries except $\sO$ and $\sO^1_+$ have either $s>0$ or $\delta>0$ and can be ignored.

It is then natural to conjecture the following one-to-one correspondence between MAD multiplets and multiplets on the Coulomb branch (where we have used the map in \eqref{mapE} and \eqref{MADcont})
\begin{equation}\label{CCmap}
\bar\CC^{\rm \, MAD}_{{p\over2},{1\over10}(2+12n+7p)(k,k-1-{p\over2})}\ \longrightarrow\ \ \bar\CC^{\rm \, Free}_{{p\over2},n+{p\over2}(k,k-1-{p\over2})}~,\ \ \ n\in\mathbb{Z}_{>1}~.
\end{equation} 
For $n=1$, the IR multiplets on the RHS above obey additional shortening conditions and become $\hat\CC$ multiplets.\footnote{For $n\le0$, the multiplets on the RHS of \eqref{CCmap} are not unitary irreducible representations and are absent. We conjecture that the corresponding multiplets on the LHS of \eqref{CCmap} are also absent (although they are not ruled out by unitarity for sufficiently large $p$).} In particular, we have
\begin{equation}\label{CChatmap}
\bar\CC^{\rm \, MAD}_{{p\over2},{7\over10}(2+p)(k,k-1-{p\over2})}\ \longrightarrow\ \ \hat\CC^{\rm \, Free}_{{p\over2}(k,k-1-{p\over2})}~.
\end{equation}
As a special case, we find, for $p=0$, UV ancestors of IR higher-spin multiplets
\begin{equation}\label{HSmap}
\bar\CC^{\rm \, MAD}_{0,{7\over5}(k,k-1)}\ \longrightarrow\ \ \hat\CC^{\rm \, Free}_{{0}(k,k-1)}~.
\end{equation}

One upshot of this discussion is that our $\bar\CC$ multiplets of interest are contained in the following OPEs
\begin{equation}
\bar\CE_{6/5}^{\times (n+p+1)}\ni\bar\CC^{\rm \, MAD}_{{p\over2},{1\over10}(2+12n+7p)(k,k-1-{p\over2})}~.
\end{equation}
This result is compatible with the conjecture in \cite{Bhargava:2022cuf} that $\bar\CE_{6/5}^{\times n}\times\CE_{-6/5}^{\times m}$ should generate the full local operator algebra of the MAD theory.

\subsec{Quasi-chiral spectrum  of the MAD theory}

Finally, combining the above discussion with the IR spectrum in \eqref{IRspectrum} we arrive at the result promised in the introduction, which we repeat here
\begin{eqnarray}\label{finalresult}
N_{\bar\CC^{\rm\, MAD}_{{p\over2},r(k,k-1-{p\over2})}}=\begin{cases}
\sum_{i=0}^p(-1)^i\Big[  N\left(p-i,n+i,k-{i\over2}\right)-N\left(p-i,n+i,k-{i\over2}-{1\over2}\right)\Big]~,\ r=r(n,p)~,\\ 
0~, \ \text{otherwise}~,
\end{cases}
\end{eqnarray}
where
\begin{eqnarray}
N(p,n,k)&:=&\sum_{ q =1}^{ q _{\rm max}}  \sum_{\ell =0}^{2k-r-p} (-1)^{ q +1} R(\ell, q +p)S(2k- q -p-\ell ,n- q +1)~,\cr r(n,p)&:=&{1\over10}(2+12n+7p) ~,\ \ \ n\in\mathbb{Z}_{\ge1}~,\ \ \ p\in\mathbb{Z}_{\ge0}~.
\end{eqnarray}

One consequence of \eqref{finalresult} is that the $Q^1_+$ and $\bar Q^2_{\dot+}$ cohomologies discussed in section \ref{cohomology} should be trivial for the MAD theory as well. This fact makes intuitive sense since the geometrical actions in \eqref{geoId1} and \eqref{geoId2} do not depend on turning on interactions (e.g., as in \eqref{Wdef}) or on the breaking of conformal symmetry. An alternative derivation of \eqref{finalresult} will be presented in Appendix~\ref{AppB}.  

Another consequence of the above discussion is that, as in \eqref{FVfactorize}, we expect a factorization of the spectrum in \eqref{finalresult}. In particular, we have
the factorization of SCPs
\begin{equation}\label{MADfactorize}
\CO_{r_{k,p} +\frac65m}=\CO_{r_{k,p}} \sO^{ m}~, \quad
\CO_r\in \bar\CC_{{p\over2},r (k,k-1-{p\over2})}~,  \quad
r_{k,p}= r(n_{k,p},p) ~,\;
n_{k,p}=2k-1-p~, \quad m \in \mathbb N~.
\end{equation}
We also therefore expect the spectrum in \eqref{finalresult} to stabilize for $r\ge r(n_{k,p},p)$. This behavior is an indication of the MAD theory's inherently Coulombic behavior. It would be interesting to understand if related behavior occurs in the rest of the quasi-chiral sector and beyond.

In the next section, we will find non-trivial confirmation of this picture from the superconformal index.

\subsec{Checks from the superconformal index}\label{index}
Here we discuss various superconformal index checks we have performed of the operator maps introduced previously. From the simple maps in \eqref{mapE} and \eqref{MADcont}, we will see that the index finds perfect agreement with the result in \eqref{finalresult}.\footnote{Although the index itself can have various cancelations, we believe this does not occur here. As we explained in the introduction, such a situation extends results on non-cancellation and minimality found in different sectors of the theory in \cite{Buican:2021elx,Bhargava:2022cuf}. Moreover, the simplicity of our inputs in \eqref{mapE} and \eqref{MADcont} combined with the robustness of the index checks suggests our picture is correct. In addition, as we will describe below, the multiplets in \eqref{finalresult} do not have cancelling index contributions amongst themselves. Finally, we will also show that the local-algebra generating conjecture in \cite{Bhargava:2022cuf} puts constraints on index cancelation.}

To that end, let us consider the superconformal index  defined by
\begin{equation}\label{indexMAD}
\CI={\rm Tr}(-1)^F
p^{j+\bar j+r}q^{-j+\bar j+r}t^{R-r} ~.
\end{equation}
As elaborated upon in \cite{Bhargava:2022cuf}, there are three classes of multiplets that contribute to the index of the MAD theory: the $\bar\CE_{6n/5}$ multiplets, the $\hat\CC_{R(j,j)}$ multiplets, and the $\bar\CC_{R,r(j,\bar j)}$ multiplets.  As a result,   the index can be written as
\begin{equation}\label{knownN0}
\CI_{\text{MAD}}=1+\CI_{\bar\CE}+\CI_{\hat\CC}+\CI_{\bar\CC}~,
\qquad 
\CI_{\alpha}=\sum_{  \bm n} N_{\alpha_{\bm n}} \mathcal I_{\alpha_{\bm n}}~,
\end{equation}
where $\CI_{\alpha}$ denotes the {\it total} index contribution from multiplets of type $\alpha$, and $\bm n$ denotes the quantum numbers of the multiplet.  The decomposition \eqref{knownN0} enables us to  isolate the index contributions of all $\CI_{\bar\CC}$ multiplets, since the rest of the contributions are known. Indeed, the multiplicities of the first two classes of multiplets are \cite{Buican:2021elx}\footnote{In \cite{Bhargava:2022cuf} we also found the exact multiplicities of the $\bar\CC_{0,r(j,0)}$ multiplets for all $r$ and $j$ (which in turn motivates a conjecture on the full local operator algebra we will return to below).}
\begin{equation}\label{knownN}
N_{\bar\CE_{6n/5}}=1~, \ \ \ \forall\ n\in\mathbb{Z}_{>0}~,\ \ \ {x^{R(R+2)}\over(1-x^2)(1-x^3)\cdots(1-x^{R+1})}=\sum_{2j=0}^{\infty}x^{2j}N_{\hat\CC_{R(j,j)}}~.
\end{equation}
Moreover, the contributions of individual short multiplets are also well known
\beqn
{\cal I}_{{\hat \CC}_{R(j, \bar j)}}&=&
(-1)^{2(j+ \bar j)}p^{j}q^{j}t^{R-j+\bar j-1}\frac{t-p q}{(1-p)(1-q)}
\left[p^{\frac12} q^{\frac12} t\chi_{j+{\frac12}}\left(\sqrt{\frac{p}{q}}\right)
-pq\chi_{j}\left(\sqrt{\frac{p}{q}}\right)\right]~,\nonumber\\
{\cal I}_{\bar\CE_{r }}&=&  p^{r-1}q^{r-1}t^{-r} \frac{(t-p)(t-q)}{(1-p)(1-q)}\ \label{indexE}~,
\nonumber\\
{\cal I}_{\bar\CC_{R,r(j, \bar j)}}&=&
-(-1)^{2(j+ \bar j)}p^{\bar j+r}q^{\bar j+r}t^{R-r-1}\frac{(t-pq)(t-p)(t-q)}{(1-p)(1-q)}\chi_{j}\Big(\sqrt{\frac{p}{q}}\Big)~,
\label{Cbarindex}
\eeqn
where $\chi_j(x):=(x^{2j+1}-x^{-2j-1})/(x-x^{-1})$ is the character for the spin $j$ irreducible representation of $SU(2)$.

Finally an expression for the index of MAD theory was obtained  in \cite{Maruyoshi:2016tqk}
 \be\label{index2}
 \cI_\text{\text{MAD}}=(p;p)(q;q) \frac{\Gamma_e\Big( (\frac{pq}{t})^{\frac65}\Big) }{\Gamma_e\Big( (\frac{pq}{t})^{\frac25}\Big)}
 \oint_C \frac{dz}{2\pi i z} \frac{ \Gamma_e\Big(  z^{\pm  }(pq)^{\frac25} t^{\frac{1}{10}}\Big) 
\Gamma_e\Big(  z^{\pm  }(pq)^{-\frac15} t^{\frac{7}{10}}\Big) 
 \Gamma_e\Big(  z^{\pm 2,0}(\frac{pq}{t})^{\frac15} \Big)}
 {2\Gamma_e (z^{\pm 2} )}~.
 \ee
This formula provides the basis for a numerical approach to compute the index. For this purpose, it is  convenient to change the variables to  $\tau$, $y$, $v$, where $p=y\tau^3$, $q=y^{-1}\tau^3$, and $t=\tau^4/v$.

We are now ready to describe our index checks. Using the decomposition in \eqref{knownN0}, we can subtract  the $\bar\CE$ and $\hat\CC$ index contributions from \eqref{index2} and  get the contributions from $\bar \cC$ multiplets:
\be\label{Iccon}
\cI_{\bar \cC}=\sum_{R,r,j,\bar j} N_{\bar \cC_{R,r(j,\bar j)}} \cI_{\bar \cC_{R,r(j,\bar j)}}
=\sum_{r,j,s} F(s,r,j) \cI_{\bar \cC_{0,r(j,\bar j=s+j  )}}~,
\ee
where $s=R+\bar j -j $, and
\be\label{FNrel}
 F(s,r,j)=\sum_{p=0}^{2s+2j } (-1)^pN_{\bar \cC_{s+p/2,r+p/2(j, s+ j -p/2)}}~.
\ee
The reason for writing the index in this form is due to multiplet recombination and, more generally, the fact that (up to a sign) the index does not distinguish between operators with quantum numbers $(R,r,j,\bar j)$ and $(R+\delta,r+\delta,j,\bar j-\delta)$. Therefore, in the decomposition of $\CI_{\bar\CC}$ into contributions from various $\bar\CC$ multiplets in \eqref{Iccon}, only $F(s,r,j)$ can be directly calculated (instead of the individual $N$ appearing in the sum).

In the case of $s=-1$, \eqref{finalresult}  enables us to count the individual $N$ summands in \eqref{FNrel}. Comparing with the $F$ computed by the index,  we have numerically checked that \eqref{FNrel} is indeed satisfied for $s=-1$.\footnote{We have numerically calculated $\cI_{\bar \cC}$ up to order $\CO(\tau^{56})$. This computation provides three-hundred and fifty-three data points with $s=-1$. Consistency with \eqref{FNrel}  therefore provides a strong check of our counting formula \eqref{finalresult}.\label{numerics}}

Note that, while the index on its own does not resolve $(R,r,j,\bar j)\to(R+\delta,r+\delta,j,\bar j-\delta)$ ambiguities, one interesting fact is that the operators in \eqref{finalresult} do not cancel amongst themselves in the MAD index (although there are infinitely many cancelations of their IR counterparts in the index on the Coulomb branch). To understand this statement, note that, independently of any ambiguities of the index, in the UV
\begin{equation}\label{UVrR}
r-R={1\over5}(1+p+6n)~.
\end{equation}
To have a cancelation, we would need to have the same $r-R$ for a fermion ($p\in\mathbb{Z}_{\rm odd}$) and  a boson ($p\in\mathbb{Z}_{\rm even}$). However, in these two cases, the parity of $p+6n$ is different and so there cannot be cancellations between fermionic and bosonic degrees of freedom in \eqref{finalresult}.\footnote{This statement means that for fixed $s,r,j$ in \eqref{FNrel}, the non-trivial terms in the sum should have the same parity and therefore identical weight factor,  $(-1)^p$. Even more strongly, we have observed numerically (up to the order in footnote \ref{numerics}) that elements of the $s=-1$ $\bar\CC$ spectrum with distinct quantum numbers correspond to distinct non-cancelling index contributions. This behavior is similar to what we observe elsewhere in the MAD spectrum: all $\bar\CE$ and $\hat\CC$ multiplets also have distinct non-cancelling index contributions for operators with distinct quantum numbers (analogous statements hold for the $s=0$ $\bar\CC$ multiplets studied in \cite{Bhargava:2022cuf}).}

\begin{figure}[h!]
\centering
\includegraphics[width=\textwidth]{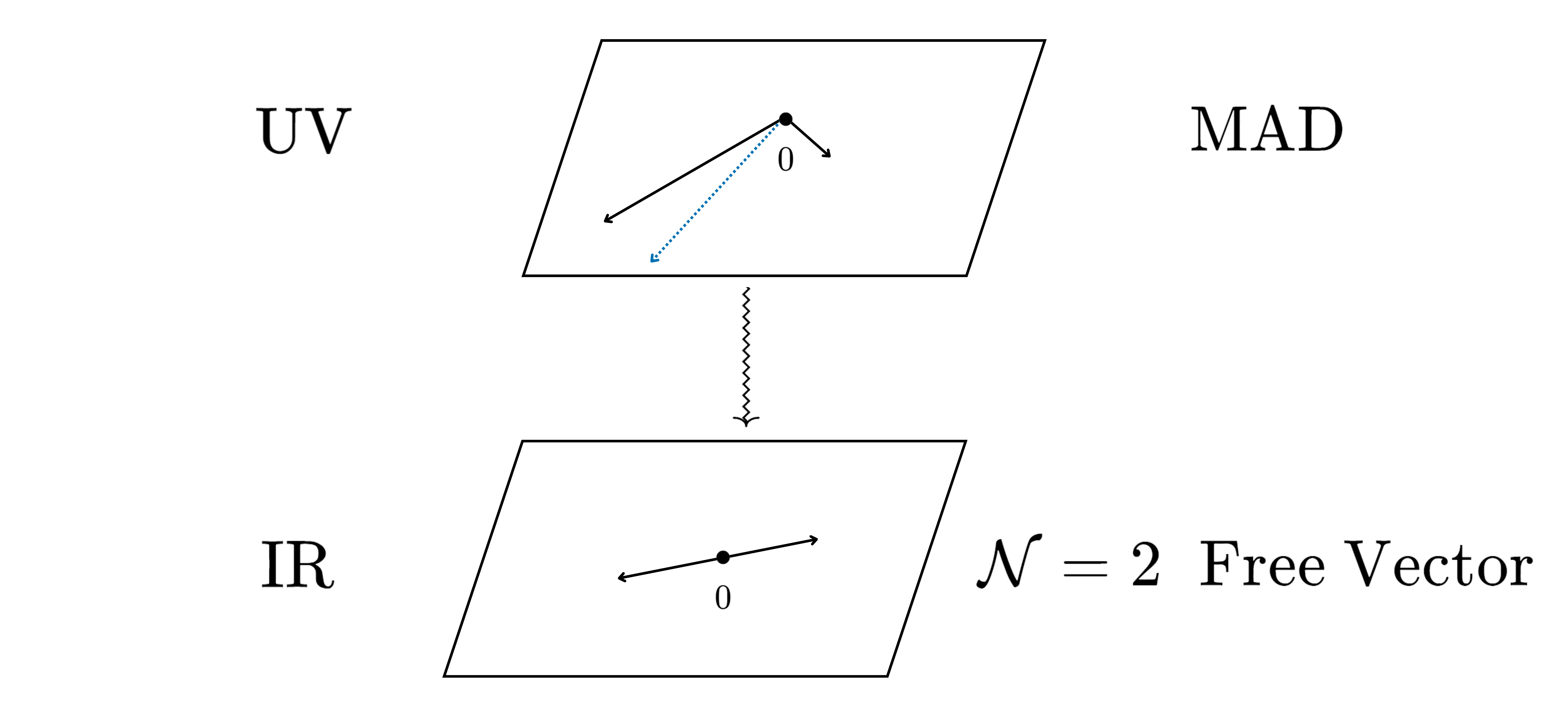}
\caption{Each complex plane contains index contributions (for some generic choice of complex fugacitites) from multiplets in the UV and IR respectively. Each multiplet contribution can be represented as a vector in $\mathbb{C}$. In the IR plane, we show cancelling index contributions for a pair of operators satisfying \eqref{IRrR} for fixed $n$ and different parity of $p$. In the UV plane, we show the contributions from the two UV avatars of these operators, where their index contributions no longer cancel (their sum is given by the blue vector) due to the constraint in \eqref{UVrR}.}\label{Flowdiag}
\end{figure}

 On the other hand, in the IR, we have
\begin{equation}\label{IRrR}
r-R=n~,
\end{equation}
and so there are infinitely many index cancellations between   even and odd $p$. The qualitative difference between the behavior in \eqref{UVrR} and \eqref{IRrR} is illustrated in figure \ref{Flowdiag}.

More generally, consider the following UV OPEs
\begin{equation}
\bar\CE_{6/5}^{\times n}\ni \tilde\sO_1\times\cdots\times\tilde\sO_n~,
\end{equation}
where the $\tilde\sO_I\in\bar\CE_{6/5}$ are any conformal primaries of the MAD $\bar\CE_{6/5}$ multiplet. Any such $\tilde\sO_I$ can be represented by the action of the $Q^i_{\alpha}$ on $\sO$. Logic similar to the one employed below \eqref{UVrR} shows that $\delta=0$ operators appearing in such OPEs with an even or odd total number of $Q^2_{\alpha}$ acting on the various $\tilde\sO_i$ cannot cancel amongst themselves within the index. Given our conjecture on the local algebra operator generation by the $\bar\CE_{6/5}^{\times n}\times\CE_{-6/5}^{\times m}$ OPEs in \cite{Bhargava:2022cuf}, it would be interesting to understand if a more detailed OPE analysis can be used to argue for the complete absence of index cancellations in the MAD theory. Conversely, finding an independent argument for the absence of index cancellations in the MAD theory would constrain the $\bar\CE_{6n/5}^{\times n}\times\CE_{-6n/5}^{\times m}$ OPEs.

\newsec{Conclusions}
In this paper, we found UV avatars of the (complex) higher-spin symmetry that emerges on the Coulomb branch of the MAD theory. In addition, we showed how these degrees of freedom are related to a rich set of operators appearing in OPEs that define a so-called quasi-chiral sector of the MAD theory. In so doing, we provided additional confirmation of our conjecture on the structure of the MAD local operator algebra in \cite{Bhargava:2022cuf}. This discussion takes us closer to computing the full protected spectrum of the MAD theory and suggests several avenues for further study:

\begin{itemize}
\item It is straightforward to complete the analysis of the quasi-chiral sector by studying UV avatars of IR $\bar\CC$ multiplets with $s_{ \CO }>-1$. This analysis should provide additional tests of our conjecture on the structure of the local operator algebra \cite{Bhargava:2022cuf}.
\item By gluing the quasi-chiral and quasi-anti-chiral sectors together, our conjecture in \cite{Bhargava:2022cuf} (supported by our analysis above) suggests that we will find the full set of local operators in the MAD theory. We expect that the $\hat\CC^{\rm MAD}_{R(j,j)}$ states (related to the 2D Lee-Yang chiral algebra \cite{Cordova:2015nma}) discussed in \cite{Buican:2021elx} will play a central role in this gluing (see the analysis in \cite{Bhargava:2022cuf} for an infinite set of $\bar\CC$ multiplets where $\hat\CC_{0(0,0)}$ plays a role; it would be interesting to generalize these multiplets and find UV avatars of the real higher-spin symmetry multiplets and their cousins). We will return to these ideas in \cite{BBJ2}.
\item We have seen that, in all classes of short multiplets studied so far, the fully refined MAD index has no cancellations. It would be interesting to understand if this is more generally true and what relation this statement has with the structure of the local operator algebra conjectured in \cite{Bhargava:2022cuf}. In fact, we have numerical evidence for a stronger behavior: namely that $s=-1$ $\bar\CC$ multiplets with different quantum numbers give different non-cancelling index contributions. As mentioned in the previous section, similar behavior occurs in all other known parts of the MAD operator spectrum. It would be interesting to prove this behavior for $s=-1$ (or else find where it breaks down) and also to understand if it holds for all $\bar\CC$ multiplets. If so, the MAD index would satisfy a sort of \lq\lq information-maximization" principle.
\item In light of the absence of MAD index cancellations discussed in the previous bullet and the fact that the index of the free vector suffers from rather generic cancelations (see the discussion around \eqref{UVrR} and \eqref{IRrR}), the map between these two observables is highly non-trivial. On the other hand, the maps between short multiplets in the MAD theory and the free vector are relatively simple. As a result, we expect the protected contributions to the thermal partition functions of these two theories to be closely related. It would be interesting to find this relation and to understand if it can be used to construct a map between the full thermal partition functions.\footnote{We thank S.~Razamat for mentioning this possibility to us.}
\item It would be interesting to extend the geometrical construction in section \ref{cohomology} to the full quasi-chiral sector and also to understand to what extent we can capture conformal symmetry (this may lead to a natural appearance of a (super)-Virasoro algebra related to that in \cite{Svrcek:2003kr}). In addition, it is natural to extend our construction to higher-dimensional Coulomb branches and to describe other multiplets like those of type $\bar\CB$ in higher-rank theories. It would then be worthwhile to make more direct contact with the Seiberg-Witten construction.
\item It would be interesting to quantitatively extend the operator factorization we found in the $s_{ \CO }=-1$ part of the MAD theory to the full theory and to understand how this interacts with the structure of the $\bar\CE_{6/5}^{\times n}\times\CE_{-6/5}^{\times m}$ OPEs. In addition, it would be interesting to investigate if there is a large charge effective field theory explanation of the operator factorization.
\item Finally, it would be useful to make contact with more general theories (e.g., see \cite{Xie:2021omd,Song:2021dhu,Kang:2022vab} for recent spectral results in some more of these theories).
\end{itemize}
We will return to some of these questions in \cite{BBJ2}.

\ack{We are grateful to Z.~Komargodski, A.~Manenti, and S.~Razamat for discussions and correspondence. M.~B. and H.~J. would also like to thank T.~Nishinaka for collaboration on related work. M.~B. would like to thank the Simons Center for Geometry and Physics and the Institute for Advanced Study for wonderful and stimulating environments during extended visits in the summer and fall, where this work was completed. In addition, M.~B. would like to thank the Aspen Center for Physics (supported by NSF grant PHY-1607611) as well as the organizers and participants of the workshop, \lq\lq Higher Symmetry and QFT" for a wonderfully insightful experience that contributed to the present paper. M.B.'s visit was partially supported by the Simons Foundation. M.~B. and H.~J. were partially supported by the grant \lq\lq Relations, Transformations, and Emergence in Quantum Field Theory" from the Royal Society and the grant \lq\lq Amplitudes, Strings and Duality" from STFC. M.~B. was also partially supported by the grant \lq\lq New Aspects of Conformal and Topological Field Theories Across Dimensions" from the Royal Society. C.~B. was partially supported by funds from Queen Mary University of London (QMUL). This research utilised QMUL's Apocrita HPC facility, supported by QMUL Research-IT. We also acknowledge the assistance of the ITS Research team at QMUL. No new data were generated or analysed during this study.}

\newpage
\begin{appendices}
\section{Various partitions of integers} \label{PartInt}
In this appendix, we introduce various kinds of partitions of integers that are used in the main body of the paper.  To that end, let $P(n,k)$ denote the  number of partitions of the integer $n$ into $k$ positive integers, and let $P(n)$ denote the  number of partitions of $n$ into  arbitrary positive integers.  These quantities can be obtained from the following generating functions
\be\label{PartPn}
\prod_{j=1}^\infty \frac{1}{1-y^j}=\sum_{m=0}^\infty P(m)y^m~,
\ee
and
\be 
\prod_{j=1}^\infty \frac{1}{1-xy^j} =\sum_{k,n=0}^\infty P(n ,k) x^k y^n~.
\ee

We are also interested in partitions into non-negative integers. Let $S(n,k)$  denote the  number of partitions of the integer $n$ into  $k$ non-negative integers. Then $S(n,k)=P(n+k,k)$. Note that we have the property $P(n)=S(n,k)$ for  $k\ge n$, as it is impossible to  partition $n$ into more than $n$ positive integers. 

In counting fermionic degrees of freedom, we are  interested in partitions into distinct integers.  To that end, let $R(n,k)$ denote the  number of  partitions of $n$ into  $k$ non-negative {\it distinct} integers. This quantity can be obtained from the following generating function
\be \label{PartRnk}
\prod_{j=1}^\infty  ({1+xy^j} )=\sum_{k,n=0}^\infty R(n-k,k) x^k y^n~.
\ee
 Using this generating function, one can prove the following recursion relations
 \beqn\label{SRrecursion}
 S(n,k)&=&S(n-k,k)+S(n,k-1)~, \\
  R(n,k)&=&R(n-k,k)+R(n-k+1,k-1)~.
 \eeqn
  
  Let us now prove some identities. For this purpose, let us  rewrite \eqref{PartRnk} as 
   \be \label{PartRnk22}
 (1+\eta xy) (1+\eta x y^2) (1+\eta xy^3) \cdots=\sum_{s,r=0}^\infty R(s,r) \eta^r x^r y^{s+r}~.
 \ee
 In addition, we have
  \be 
 \frac{1}{(1-xy) (1-x y^2) (1-xy^3) \cdots} 
=\sum_{k,n=0}^\infty S(n-k ,k) x^k y^n~.
 \ee
 
Taking  the product of the two generating functions above and setting $\eta=-1$, we find
 \beqn
 1&=&
 \sum_{s,r=0}^\infty \sum_{k,n=0}^\infty S(n-k ,k)   R(s,r) (-1)^r  x^{k+r}y^{n+s+r}
 \\&=&
 \sum_{s,r=0}^\infty \sum_{k,n=0}^\infty (S(n,k)-S(n  ,k-1))   R(s,r) (-1)^r  x^{k+r}y^{n+s+r}
 \\&=& 
 \sum_{s,r=0}^\infty \sum_{k,n=0}^\infty (S(n-r-s,k-r)-S(n -r-s ,k-r-1))   R(s,r) (-1)^r  x^{k }y^{n} ~, \qquad
\eeqn
where we used the recursion relation \eqref{SRrecursion} in the second equality. These manipulations mean that
\beqn
 \sum_{s,r=0}^\infty  (S(n-r-s,k-r)-S(n -r-s ,k-r-1))   R(s,r) (-1)^r   
=\delta_{k,0}\delta_{n,0}~.
 \eeqn
 We can then sum over $k$, and find that
 
      \be \label{RSeqs}
 \sum_{s,r=0}^\infty   S(n-r-s,k-r)   R(s,r) (-1)^r   
= \delta_{n,0}~, \qquad k\ge 0~.
 \ee 
 When $k\ge n$, we have $ S(n-r-s,k-r)  =P(n-r-s)$, then the above identity reduces to
 \be
\sum_{ s, r=0}^\infty P(n-r-s) R(s,r) (-1)^r=\delta_{n,0}~,
\ee
or equivalently
\be\label{idPart}
\sum_{ r=1}^\infty \sum_{ s=0  }^\infty  (-1)^{r+1}P(n-r-s) R(s,r) =P(n)-\delta_{n,0}~.
\ee

\newsec {An alternate method for counting  $\bar {\mathcal C}$ multiplets} \label{AppB}

In this appendix, we present an alternate (and in some senses more direct) approach to counting the multiplicity of  $\bar {\mathcal C}_{R,r,(j,\bar j)}$ multiplets satisfying $s=-1$. This approach is also illuminating and therefore worth elaborating on.
 
Following our prescription in the main text we count $\bar\CC$ multiplets via their level-one descendants, which have much simpler representations
\beqn\label{Rrnkl}
 &&\CO \in \bar {\mathcal C}_{R,r,(j,\bar j)} \text{ with } R+\bar j -j+1=0
 \\& \longleftrightarrow  &
X=\bar Q^1_{\dot +}\CO \sim \lambda^l\phi^k\p^n\in {\mathcal G}_{n,k,l} ~,
\label{Xops}
\eeqn
where ${\mathcal G}_{n,k,l} $ is the set of operators with the prescribed  number of fields and derivatives (schematically indicated by \lq\lq$\lambda^{l}\phi^k\partial^n$"), namely 
\be\label{Gnkl}
{\mathcal G}_{n,k,l} :=\Big\{\sum_{\bm a, \bm b} C_{\bm a, \bm b} \p^{a_1}_{+\dot+} \lambda_+^1\cdots
\p^{a_l}_{+\dot+} \lambda_+^1
\p^{b_1}_{+\dot+} \phi \cdots
\p^{b_k}_{+\dot+} \phi
\Big| \sum_i  a_i+\sum_j b_j =n
\Big\}~.
\ee
For notational simplicity in what follows, we use  $\lambda^l\phi^k\p^n$ to denote an arbitrary operator of the form specified above. 

The numbers $n,k,l$ in   \eqref{Xops} are determined by the charges 
\beqn\label{chargesNumb}
 n=2j-2R-1~, \quad k=-1-2R-\frac{R+r+1}{r_0}~,\quad l=2R+1~, \quad \bar j = j-R-1~,
\eeqn
where $r$ is the $U(1)_r$ charge/scaling dimension of Coulomb branch operator $\phi$. The main text justifies the idea that we can consider either $\phi$ and $\lambda^1_+$ (the free vector degrees of freedom) or $\sO$ and $\sO_+$ (the MAD degrees of freedom). Therefore, while we will use the notation $\phi$ and $\lambda^1_+$ here (with $r_0=1$ for the primary), we can think of our manipulations below as applying equally well to $\sO$ and $\sO_+$ (with $r_0=6/5$ for the primary).\footnote{A priori, one might have thought that null relations in the MAD theory could prevent such a naive substitution. However, the results in the main text show this is not the case.}

 Since $X$ is the $\bar Q^1_{\dot +} $ descendant of the SCP $\CO$, we expect that $X$ is annihilated by all the special supercharges $S, \bar S$ except for $ \bar S^2_{\dot -} $, so 
 \be
 \bar S^2_{\dot -} X\neq 0\qquad \longleftrightarrow \qquad X= \bar Q^1_{\dot +}  \CO~. 
 \ee
 From the explicit general form of $X$,  we easily check that 
  \be
 [  S^i_{+} ,X] =[\bar S^i_{\dot +},X]=[  S^1_{-},X]=0~.
 \ee
 So the  only  non-trivial conditions are
 \be\label{SXcond}
 [S^2_-,X]=[\bar S^1_{\dot -},X]=0~,
\ee
or, equivalently,
 \beqn\label{S2X}
 [S^2_-,X]= 0\qquad &&\longleftrightarrow \qquad X\neq Q^1_+X' ~,\qquad \forall X~,
\\ \label{S1X}
[\bar S^1_{\dot -},X]= 0\qquad&& \longleftrightarrow \qquad X\neq \bar Q^2_{\dot+}X'', \qquad \forall X''~.
 \eeqn

Before imposing these constraints, we can count the number of linearly independent operators 
\be
\CN(n,k,l)=|   {\mathcal G}_{n,k,l}  |=\sum_{m=0}^n S(n-m, k) R(m,l)~.
\ee
where \lq\lq$|\cdots |$" denotes the dimension of the enclosed space.
 
For later purposes, we also introduce the following notation
\be
X_{i,j}\sim \lambda^{l-j+i}\phi^{k-i+j}\p^{n-i} \in {\mathcal G}_{n-i,k-i+j,l-j+i}~,
\ee
and we identify $X=X_{0,0}\sim \lambda^l\phi^k\p^n$.

Now, since
\be\label{Qbaraction}
\bar Q_{\dot+}^2 \p^a \lambda\sim \p^{a+1}\phi~, \qquad
\bar Q_{\dot+}^2 \p^a \phi=0~,
\ee
\be\label{Qaction}
 Q_{ +}^1 \p^a \phi\sim\p^a \lambda~, \qquad  Q_{ +}^1 \p^a \lambda=0~,
\ee
we find  
\beqn
&&\bar Q_{\dot+}^2:{\mathcal G}_{n-i,k-i+j,l-j+i}\to {\mathcal G}_{n-i+1,k-i+j+1,l-j+i-1}~, \quad X_{i,j}\mapsto X_{i-1,j}~,
\\&&
 Q_{ +}^1: {\mathcal G}_{n-i,k-i+j,l-j+i}\to {\mathcal G}_{n-i,k-i+j-1,l-j+i+1}~, \qquad
 X_{i,j}\mapsto X_{i ,j-1}~.
\eeqn

As a result, we find the double complex in figure \ref{doublecomplex}. As we showed in subsection \ref{cohomology}, the cohomology is trivial.  Therefore, whenever $ \bar Q_{\dot+}^2 Y=0$, we have $Y=\bar Q_{\dot+}^2 Z$, and similarly for $ Q_{ +}^1$. Due to the non-trivial anti-commutation relation $\{Q_{ +}^1,\bar Q_{\dot+}^2\}=\p_{+\dot+}$, the  diagram 
looks non-commutative. However, one can make the diagram commutative by regarding $X$ as an equivalence class, namely $X\sim X'$ if $X=\pm X'+\p_{+\dot+} X''$. 

\begin{figure}[h]
\centering
\begin{tikzcd}
 &   \vdots\arrow[d, " Q_{ +}^1"] &  \vdots \arrow[d, " Q_{ +}^1"]
 \\
  \cdots \arrow[r, "\bar Q_{\dot+}^2"] &  {\mathcal G}_{n-1,k,l}\arrow[d, " Q_{ +}^1"] \arrow[r, "\bar Q_{\dot+}^2"] &    {\mathcal G}_{n,k+1,l-1}\arrow[d, " Q_{ +}^1"]
  \\
   \cdots \arrow[r,  "\bar Q_{\dot+}^2"]  &  {\mathcal G}_{n-1,k-1,l+1}\arrow[r, "\bar Q_{\dot+}^2"] & {\mathcal G}_{n,k,l}
\end{tikzcd}
\caption{Double complex formed by quasi-chiral operators with $s=-1$.}\label{doublecomplex}
\end{figure}
 
 Now we would like to count $X$ operators as descendants of $\CO$ described in \eqref{Rrnkl}, subject to the constraints in \eqref{S2X}\eqref{S1X}
 \beqn
\CN' (n,k,l) &=&
|\{X| QX \neq 0 , \bar QX\neq 0\}|
\\&=&
|\{ X  \}| - |\{X | QX=0\}| - |\{X | \bar QX=0\} |+ |\{X | QX=\bar QX=0\}|
\\&=&
|\{ X \}| - |\{X_{0,1}\}| -|\{X_{1,0}\}+\{X_{1,1}\}|
\\&&
+|\{X_{0,1}| QX_{0,1}=0\}|+|\{X_{1,0}|   \bar QX_{1,0}=0\}|
\\&&
-|\{X_{1,1}|     QX_{1,1}=0\} | -|\{X_{1,1}|  \bar  QX_{1,1}=0\} | +|\{X_{1,1}| QX_{1,1}= \bar  QX_{1,1}=0\} | ~,
\qquad\quad
\eeqn
where $Q:=Q_+^1, \bar Q:=\bar Q_{\dot+}^2,X\equiv X_{0,0}\in {\mathcal G}_{n,k,l}, X_{i,j}\in {\mathcal G}_{n-i,k-i+j,l-j+i}$, and 
\be
|\{X_{i,j}\}|=|{\mathcal G}_{n-i,k-i+j,l-j+i}|=\CN(n-i,k-i+j,l-j+i)~.
\ee
In the derivation above, we have used the following relation
\beqn
|\{X | QX=0\}|&=&|\{X |  X=QX_{0,1}\} |=|\{X_{0,1}\}|-|\{X_{0,1}|   QX_{0,1}=0\} |~,
\\
|\{X | \bar QX=0\}|&=&|\{X |  X=QX_{0,1}\} |=|\{X_{0,1}\}|-|\{X_{0,1}|   QX_{0,1}=0\} |~,\qquad
\\
|\{X | QX=\bar QX=0\}|&=& |\{X | X=Q\bar QX_{1,1}\}| =|\{X_{1,1}\}|-|\{X_{1,1}| Q\bar  QX_{1,1}=0\} | 
\nonumber\\&=&
|\{X_{1,1}\}|-|\{X_{1,1}| QX_{1,1}= 0\} | 
-|\{X_{1,1}|  \bar  QX_{1,1}=0\} | 
\nonumber \\& &+
|\{X_{1,1}| QX_{1,1}= \bar  QX_{1,1}=0\} | ~.
\eeqn

By induction, we then find
\beqn
\CN '(n,k,l) &=&\sum_{i,j=0}^\infty (-1)^{i+j} |\{X_{ij} \in {\mathcal G}_{n-i,k-i+j,l-j+i}\}|
=\sum_{i,j=0}^\infty (-1)^{i+j}\CN(n-i,k-i+j,l-j+i)
\nonumber\\&=&
\sum_{i,j,m=0}^\infty (-1)^{i+j}S(n-i-m, k-i+j) R(m,l+i-j)~.
\label{Nx3}
\eeqn
Finally, we want to count primary operators, so we need to subtract the conformal descendants 
\be\label{primaryCN}
\CN_{\bar \CC_{R,r(j,\bar j)}}=
\CN ' (n,k,l) -\CN ' (n- 1 ,k,l) ~,
\ee
where the weight in the multiplet is determined by \eqref{chargesNumb}. This is the main result of this appendix, enabling us to count the    $\bar\CC_{R,r(j,\bar j)}$ multiplets with $R+\bar j -j +1=0$.

We would like to contrast  the results here with \eqref{subtractpGen}, which in the current notation  reduces to 
\beqn\label{Nprime2}
 N' &=&
\sum_{i,  m=0}^\infty \sum_{j=0}^{l-1}(-1)^{i+j}S(n-i-m, k-i+j) R(m,l+i-j)~.
\eeqn
The only difference between \eqref{Nx3} and \eqref{Nprime2} is  the range of $j$. This difference turns out to be inconsequential. More specifically
\beqn
\CN'-N '&=&\sum_{i,  m=0}^\infty \sum_{j=l}^{\infty}(-1)^{i+j}S(n-i-m, k-i+j) R(m,l+i-j)
\\&=&
\sum_{i,  m=0}^\infty \sum_{j=0}^{\infty}(-1)^{i-j+l}S(n-i-m, k+l-i+j) R(m, i-j)
\\&=&(-1)^l
\sum_{j=0}^{\infty}\sum_{r,  m=0}^\infty  (-1)^{r}S(n-j-r-m, k+l-r) R(m,r)
\\&=&(-1)^l
\sum_{j=0}^{\infty}\delta_{n,j} 
\\&=&(-1)^l~,
\eeqn
where we use the identity  \eqref{RSeqs} in the fourth equality above.  
This simple difference cancels out when considering the primaries \eqref{primaryCN}, so the two methods give the same result $\CN_{\bar\cC}=N_{\bar\cC}$, except for $n=0$ or, equivalently, the $R=j+1/2$ case. In this situation, the $X$ operators have no derivatives and thus  are always conformal primary. However, these operators don't belong to  $\bar \cC$ multiplets.  

At the level of non-primaries, the difference  between \eqref{Nx3} and \eqref{Nprime2} can be understood physically as follows. For simplicity, let us consider the case of $l=1$ and $R=0$, namely there is only one gaugino in $X$. There is an operator $X\propto \p^n(   \phi^k \lambda )\propto Q_+^1(\p^n \phi^{k+1})$. Therefore, as a $Q_+^1$ descendant, it is subtracted in the counting of  \eqref{Nx3}, but not in   \eqref{Nprime2}. However, when considering primaries, it is also subtracted in  \eqref{Nprime2} as it is a conformal descendant. 
\end{appendices}

\newpage

\bibliography{chetdocbib}

\begin{thebibliography}{10}
\ifx\href\asklfhas\newcommand{\href}[2]{#2}\fi
\ifx\arxivref\asklfhas\newcommand{\arxivref}[2]{\href{http://arxiv.org/abs/#1}{#2}}\fi
\ifx\doiref\asklfhas\newcommand{\doiref}[2]{\href{http://dx.doi.org/#1}{#2}}\fi
\parskip 0pt
\normalsize

\bibitem{Seiberg:1994rs}
N.~Seiberg \& E.~Witten,
\textit{``{Electric - magnetic duality, monopole condensation, and confinement
  in N=2 supersymmetric Yang-Mills theory}''},
\doiref{10.1016/0550-3213(94)90124-4}{Nucl.~Phys.~B \textbf{426}, 19
  (1994)\ignorespaces}\ignorespaces,
\normalsize{\texttt{\arxivref{hep-th/9407087}{hep-th/9407087}}}\ignorespaces,
[Erratum: Nucl.Phys.B 430, 485--486 (1994)]\ignorespaces
\bibitem{Seiberg:1994aj}
N.~Seiberg \& E.~Witten,
\textit{``{Monopoles, duality and chiral symmetry breaking in N=2
  supersymmetric QCD}''},
\doiref{10.1016/0550-3213(94)90214-3}{Nucl.~Phys.~B \textbf{431}, 484
  (1994)\ignorespaces}\ignorespaces,
\normalsize{\texttt{\arxivref{hep-th/9408099}{hep-th/9408099}}}\ignorespaces
\bibitem{Beisert:2010jr}
N.~Beisert et~al.,
\textit{``{Review of AdS/CFT Integrability: An Overview}''},
\doiref{10.1007/s11005-011-0529-2}{Lett.~Math.~Phys. \textbf{99}, 3
  (2012)\ignorespaces}\ignorespaces,
\normalsize{\texttt{\arxivref{1012.3982}{arXiv:1012.3982
  \![hep-th]}}}\ignorespaces
\bibitem{Gadde:2010zi}
A.~Gadde, E.~Pomoni \& L.~Rastelli,
\textit{``{Spin Chains in $\mathcal{N}$=2 Superconformal Theories: From the
  $\mathbb{Z}_{2}$ Quiver to Superconformal QCD}''},
\doiref{10.1007/JHEP06(2012)107}{JHEP \textbf{1206}, 107
  (2012)\ignorespaces}\ignorespaces,
\normalsize{\texttt{\arxivref{1006.0015}{arXiv:1006.0015
  \![hep-th]}}}\ignorespaces
\bibitem{Arkani-Hamed:2022rwr}
N.~Arkani-Hamed, L.~J. Dixon, A.~J. McLeod, M.~Spradlin, J.~Trnka \&
  A.~Volovich,
\textit{``{Solving Scattering in $N$ = 4 Super-Yang-Mills Theory}''},
\normalsize{\texttt{\arxivref{2207.10636}{arXiv:2207.10636
  \![hep-th]}}}\ignorespaces
\bibitem{Coleman:1967ad}
S.~R. Coleman \& J.~Mandula,
\textit{``{All Possible Symmetries of the S Matrix}''},
\doiref{10.1103/PhysRev.159.1251}{Phys.~Rev. \textbf{159}, 1251
  (1967)\ignorespaces}\ignorespaces
\bibitem{Haag:1974qh}
R.~Haag, J.~T. Lopuszanski \& M.~Sohnius,
\textit{``{All Possible Generators of Supersymmetries of the s Matrix}''},
\doiref{10.1016/0550-3213(75)90279-5}{Nucl.~Phys.~B \textbf{88}, 257
  (1975)\ignorespaces}\ignorespaces
\bibitem{Maldacena:2011jn}
J.~Maldacena \& A.~Zhiboedov,
\textit{``{Constraining Conformal Field Theories with A Higher Spin
  Symmetry}''},
\doiref{10.1088/1751-8113/46/21/214011}{J.~Phys.~A \textbf{46}, 214011
  (2013)\ignorespaces}\ignorespaces,
\normalsize{\texttt{\arxivref{1112.1016}{arXiv:1112.1016
  \![hep-th]}}}\ignorespaces
\bibitem{Alba:2015upa}
V.~Alba \& K.~Diab,
\textit{``{Constraining conformal field theories with a higher spin symmetry in
  $d > 3$ dimensions}''},
\doiref{10.1007/JHEP03(2016)044}{JHEP \textbf{1603}, 044
  (2016)\ignorespaces}\ignorespaces,
\normalsize{\texttt{\arxivref{1510.02535}{arXiv:1510.02535
  \![hep-th]}}}\ignorespaces
\bibitem{Aasen:2020jwb}
D.~Aasen, P.~Fendley \& R.~S.~K. Mong,
\textit{``{Topological Defects on the Lattice: Dualities and Degeneracies}''},
\normalsize{\texttt{\arxivref{2008.08598}{arXiv:2008.08598
  \![cond-mat.stat-mech]}}}\ignorespaces
\bibitem{Argyres:1995jj}
P.~C. Argyres \& M.~R. Douglas,
\textit{``{New phenomena in SU(3) supersymmetric gauge theory}''},
\doiref{10.1016/0550-3213(95)00281-V}{Nucl.~Phys.~B \textbf{448}, 93
  (1995)\ignorespaces}\ignorespaces,
\normalsize{\texttt{\arxivref{hep-th/9505062}{hep-th/9505062}}}\ignorespaces
\bibitem{Argyres:1995xn}
P.~C. Argyres, M.~R. Plesser, N.~Seiberg \& E.~Witten,
\textit{``{New N=2 superconformal field theories in four-dimensions}''},
\doiref{10.1016/0550-3213(95)00671-0}{Nucl.~Phys.~B \textbf{461}, 71
  (1996)\ignorespaces}\ignorespaces,
\normalsize{\texttt{\arxivref{hep-th/9511154}{hep-th/9511154}}}\ignorespaces
\bibitem{Argyres:2015ffa}
P.~Argyres, M.~Lotito, Y.~L\"u \& M.~Martone,
\textit{``{Geometric constraints on the space of $ \mathcal{N} $ = 2 SCFTs.
  Part I: physical constraints on relevant deformations}''},
\doiref{10.1007/JHEP02(2018)001}{JHEP \textbf{1802}, 001
  (2018)\ignorespaces}\ignorespaces,
\normalsize{\texttt{\arxivref{1505.04814}{arXiv:1505.04814
  \![hep-th]}}}\ignorespaces
\bibitem{Buican:2021elx}
M.~Buican, H.~Jiang \& T.~Nishinaka,
\textit{``{Spin Thresholds, RG Flows, and Minimality in 4D $\mathcal{N}=2$
  QFT}''},
\normalsize{\texttt{\arxivref{2112.05925}{arXiv:2112.05925
  \![hep-th]}}}\ignorespaces
\bibitem{Bhargava:2022cuf}
C.~Bhargava, M.~Buican \& H.~Jiang,
\textit{``{On the protected spectrum of the minimal Argyres-Douglas theory}''},
\doiref{10.1007/JHEP08(2022)132}{JHEP \textbf{2208}, 132
  (2022)\ignorespaces}\ignorespaces,
\normalsize{\texttt{\arxivref{2205.07930}{arXiv:2205.07930
  \![hep-th]}}}\ignorespaces
\bibitem{Dolan:2002zh}
F.~A. Dolan \& H.~Osborn,
\textit{``{On short and semi-short representations for four-dimensional
  superconformal symmetry}''},
\doiref{10.1016/S0003-4916(03)00074-5}{Annals~Phys. \textbf{307}, 41
  (2003)\ignorespaces}\ignorespaces,
\normalsize{\texttt{\arxivref{hep-th/0209056}{hep-th/0209056}}}\ignorespaces
\bibitem{Dobrev:1985qv}
V.~K. Dobrev \& V.~B. Petkova,
\textit{``{All Positive Energy Unitary Irreducible Representations of Extended
  Conformal Supersymmetry}''},
\doiref{10.1016/0370-2693(85)91073-1}{Phys.~Lett.~B \textbf{162}, 127
  (1985)\ignorespaces}\ignorespaces
\bibitem{Cordova:2016emh}
C.~Cordova, T.~T. Dumitrescu \& K.~Intriligator,
\textit{``{Multiplets of Superconformal Symmetry in Diverse Dimensions}''},
\doiref{10.1007/JHEP03(2019)163}{JHEP \textbf{1903}, 163
  (2019)\ignorespaces}\ignorespaces,
\normalsize{\texttt{\arxivref{1612.00809}{arXiv:1612.00809
  \![hep-th]}}}\ignorespaces
\bibitem{BBJ2}
C.~Bhargava, M.~Buican \& H.~Jiang,
\textit{``{The Full Protected Spectrum of the Minimal Argyres-Douglas Theory
  (in progress)}''}
\bibitem{Maruyoshi:2016tqk}
K.~Maruyoshi \& J.~Song,
\textit{``{Enhancement of Supersymmetry via Renormalization Group Flow and the
  Superconformal Index}''},
\doiref{10.1103/PhysRevLett.118.151602}{Phys.~Rev.~Lett. \textbf{118}, 151602
  (2017)\ignorespaces}\ignorespaces,
\normalsize{\texttt{\arxivref{1606.05632}{arXiv:1606.05632
  \![hep-th]}}}\ignorespaces
\bibitem{Beem:2013sza}
C.~Beem, M.~Lemos, P.~Liendo, W.~Peelaers, L.~Rastelli \& B.~C. van~Rees,
\textit{``{Infinite Chiral Symmetry in Four Dimensions}''},
\doiref{10.1007/s00220-014-2272-x}{Commun.~Math.~Phys. \textbf{336}, 1359
  (2015)\ignorespaces}\ignorespaces,
\normalsize{\texttt{\arxivref{1312.5344}{arXiv:1312.5344
  \![hep-th]}}}\ignorespaces
\bibitem{Gadde:2011uv}
A.~Gadde, L.~Rastelli, S.~S. Razamat \& W.~Yan,
\textit{``{Gauge Theories and Macdonald Polynomials}''},
\doiref{10.1007/s00220-012-1607-8}{Commun.~Math.~Phys. \textbf{319}, 147
  (2013)\ignorespaces}\ignorespaces,
\normalsize{\texttt{\arxivref{1110.3740}{arXiv:1110.3740
  \![hep-th]}}}\ignorespaces
\bibitem{Maldacena:2012sf}
J.~Maldacena \& A.~Zhiboedov,
\textit{``{Constraining conformal field theories with a slightly broken higher
  spin symmetry}''},
\doiref{10.1088/0264-9381/30/10/104003}{Class.~Quant.~Grav. \textbf{30}, 104003
  (2013)\ignorespaces}\ignorespaces,
\normalsize{\texttt{\arxivref{1204.3882}{arXiv:1204.3882
  \![hep-th]}}}\ignorespaces
\bibitem{Cordova:2015nma}
C.~Cordova \& S.-H. Shao,
\textit{``{Schur Indices, BPS Particles, and Argyres-Douglas Theories}''},
\doiref{10.1007/JHEP01(2016)040}{JHEP \textbf{1601}, 040
  (2016)\ignorespaces}\ignorespaces,
\normalsize{\texttt{\arxivref{1506.00265}{arXiv:1506.00265
  \![hep-th]}}}\ignorespaces
\bibitem{Svrcek:2003kr}
P.~Svrcek,
\textit{``{On nonperturbative exactness of Konishi anomaly and the
  Dijkgraaf-Vafa conjecture}''},
\doiref{10.1088/1126-6708/2004/10/028}{JHEP \textbf{0410}, 028
  (2004)\ignorespaces}\ignorespaces,
\normalsize{\texttt{\arxivref{hep-th/0311238}{hep-th/0311238}}}\ignorespaces
\bibitem{Xie:2021omd}
D.~Xie \& W.~Yan,
\textit{``{A study of N =1 SCFT derived from N =2 SCFT: index and chiral
  ring}''},
\normalsize{\texttt{\arxivref{2109.04090}{arXiv:2109.04090
  \![hep-th]}}}\ignorespaces
\bibitem{Song:2021dhu}
J.~Song,
\textit{``{Vanishing short multiplets in rank one 4d/5d SCFTs}''},
\normalsize{\texttt{\arxivref{2109.05588}{arXiv:2109.05588
  \![hep-th]}}}\ignorespaces
\bibitem{Kang:2022vab}
M.~J. Kang, C.~Lawrie, K.-H. Lee \& J.~Song,
\textit{``{Operator spectroscopy for 4d SCFTs with a=c}''},
\normalsize{\texttt{\arxivref{2210.06497}{arXiv:2210.06497
  \![hep-th]}}}\ignorespaces
\end{thebibliography}

\end{document}